\documentclass[aps,prd,nofootinbib]{revtex4-1}
\usepackage{graphicx,soul}
\usepackage{subfigure}
\usepackage{epsfig}
\usepackage{amstext,amsmath,amsfonts,amssymb,amsthm}
\usepackage{color}

\usepackage{verbatim}
\usepackage{lipsum}
\usepackage{color,array,wrapfig}
\usepackage{mathtools}

\usepackage{xcolor}
\usepackage[normalem]{ulem}

\def\eq#1{{Eq.~(\ref{#1})}}

\def\fig#1{{Fig.~\ref{#1}}}

\def\be{\begin{equation}}
\def\ee{\end{equation}}
\def\bes{\begin{eqnarray}}
\def\ees{\end{eqnarray}}
\def\ba{\begin{align}}
\def\ea{\end{align}}
\def\bwt{\begin{widetext}}
\def\ewt{\end{widetext}}

\def\aa{\alpha}
\def\tt{\tau}

\def\yy{w}
\def\aa{\aleph}
\def\tt{\overline{T}}
\def\kp{k_p}

\newbox\one
\newbox\two
\long\def\loremlines#1{%
    \setbox\one=\vbox {%
       Test.\footnote{a footnote}%
      \lipsum\footnote{Another footnote.}%
     }
   \setbox\two=\vsplit\one to #1\baselineskip
   \unvbox\two}
\begin{document}
\title{Gravitational wave generation in a viable
scenario of inflationary magnetogenesis}

\author{Ramkishor Sharma$^{1}$}
\email{ramkishor@iucaa.in}
\author{Kandaswamy Subramanian$^{1}$} 
\email{kandu@iucaa.in}
\author{T. R. Seshadri$^{2}$}
\email{trs@physics.du.ac.in} 
\affiliation{$^{1}$ IUCAA, Post Bag 4, Pune University Campus, Ganeshkhind, Pune$-$411007 India.}
\affiliation{$^{2}$Department of Physics \& Astrophysics, University of Delhi, New Delhi$-$110007 India.}
\begin{abstract}
Generation of magnetic fields during inflation is a promising mechanism 
for the origin of the
observed large scale magnetic fields
in the universe. Among several attempts, a popular model is
one
where the inflaton and the electromagnetic field are coupled through a coupling function $f$ leading to a term in the Lagrangian density of the form, $f^2 F^{\mu \nu} F_{\mu\nu}$.
A number of potential difficulties with such models have been raised in the literature.
In our earlier work, we have suggested viable models of inflationary
magnetogenesis which
avoid these problems and at the same time can lead to either nonhelical or helical magnetic fields
of astrophysical interest. Our models require a low energy scale for inflation and reheating 
(reheating temperature, $T_R < 10^4$ GeV)
and generate a blue spectrum of electromagnetic (EM) field which peaks around the horizon
scale of reheating. 
We show here that the anisotropic stress associated with these EM
fields naturally
source the production of a
stochastic background of Gravitational waves (GW) with frequencies in the range of tens of nano Hertz to milli Hertz. These two extremes of the range can be probed respectively by
pulsar timing arrays (PTA) experiments and the upcoming Laser Interferometric Space Array (LISA).
The peak value of the GW spectrum energy represented by $d \Omega_{GW}/d \ln k$ is $ 10^{-6} $ for the models which lead to
nonhelical primordial fields
and $2 \times 10^{-6} $ for the helical case 
for $T_R=100$ GeV. 
In this case the spectrum peaks at a frequency $30 \mu$Hz for non helical case and at $40 \mu$Hz for helical case. These values are obtained when the ratio of EM energy density to the cosmological density at reheating
$\epsilon \sim 1$ and decrease approximately as $\epsilon^2$ for smaller values.
The amplitude is similar for a lower value of $T_R$, but the frequency
at which the GW spectrum peaks decreases as $T_R$.
The gravitational waves generated are unpolarized if the EM fields are nonhelical but
are circularly polarised for helical primordial fields. 
If detected in future these gravitational waves will 
provide a unique probe of such models of inflationary magnetogenesis.
\end{abstract}

\maketitle
\section{Introduction}

The discovery of gravitational waves by LIGO and VIRGO detectors from binary black hole and 
neutron star black hole binary mergers
\cite{abbott2016f,abbott2016s,abbott2017f,abbott2017s,abbott2017t,abbott2017f} 
opened a new era in astronomy. Gravitational waves (GW) can even probe sources which are not detectable through electromagnetic radiation like black hole mergers. Primordial GW can be used to 
probe various epochs in the early Universe. One of these epochs is the 
inflationary era during which the universe underwent a rapid accelerated expansion. The inflationary framework provides a solution to several problems in standard cosmology like horizon and flatness problems \cite{guth1980}. It
also gives a natural explanation for the origin of initial density fluctuations \cite{guth1982,bardeen1983} which 
are later amplified via gravity to form large-scale structures
in the universe. Tensor perturbations (gravitational waves) are also produced in a manner similar to that of scalar density perturbations during inflation \cite{rubakov1982,starobinsky1983}. These tensor perturbations travel freely after generation as their interaction with the rest of the fluid is very weak. Since, the energy scale at which inflation took place is not known, the present observations only put an upper bound on this scale of inflation from the non-detection of tensor perturbations in the cosmic microwave background radiation \cite{planck2018inflation}. 

There are various other epochs in the early universe where the production of gravitational waves (GW) could have taken place. These include the production of GW from braneworlds \cite{randall2006, durrer2007}, topological defects \cite{durrer1998}, phase transitions \cite{hogan1986, kosowsky1991, kosowsky1992, kosowsky1992a, kamionkowski1993, caprini2007, huber2008, caprini2009, kosowsky2001, dolgov2002, caprini2006, gogoberidze2007, megevand2008, kahniashvili2008, caprini2001, kahniashvili2009, apreda2001, nicolis2003, grojean2006} and primordial turbulence \cite{tinagw2008,tinagw2007,tinagw2005,anand2018}. Gravitational waves may be represented by the transverse traceless (TT) part of the metric perturbations. They are sourced by the corresponding TT part of the Energy momentum tensor. Indeed any process which generates an anisotropic stress can produce GW. This can happen, for example, if magnetic fields are generated during phase transition or during inflation.

In this paper, we focus on the production of the gravitational waves from the primordial magnetic fields which are generated during inflation. Magnetic fields have been observed over a wide range of scales in the universe \cite{r-beck,clarke,widrow,neronov,taylor2011}. These fields are assumed to be generated by the amplification of seed fields via flux freezing evolution followed by a turbulent dynamo mechanism \cite{kandu2019}. 
A number of scenarios of generation of 
seed magnetic fields have been suggested in literature such as generation during inflation \cite{turner-widrow, ratra, takahashi2005, shiv2005, martin-yokoyama, campanelli2008, rajeev2010, agullo2013,rajeev2013,caprini2014,kobayashi2014, atmjeet2014, atmjeet2015,sriram2015, Campanelli2015, 1475-7516-2015-03-040,arun2015,sriram2016,fujita2016,shinji2016,sumanta2018,fujita2019}, phase transitions \cite{vachaspati, Sigl:1996dm, kisslinger,qcd}, recombination, reionization 
and structure formation \cite{biermann,fenu,kandu1994,Zweibel2000,kulsrud1997}. 
The importance of inflationary scenariois of magnetic field generation as against other mechanisms lies in the fact that the former gives a natural way of generating fields coherent on large length scales. A popular model for such generation is one where one couples a time dependent function to the usual 
electromagnetic (EM) action. In particular \citet{ratra} 
model takes the lagarangian density of the form $f^2 F^{\mu \nu} F_{\mu \nu}$ where $f$ is a function of inflaton field and $F_{\mu \nu}$ the electromagnetic field tensor. 
Although this model generates magnetic fields of sufficient strength to satisfy a number of observational constraints, 
it suffers from the back-reaction and strong coupling problems 
\cite{mukhanov2009}. Another potential difficulty for such magnetogenesis
scenarios arises due to charged particle production
by the Schwinger mechanism which arrests the growth of magnetic field \cite{kobayashi:2014}.

In a recent study by \citet{sharma2017},
we have suggested a scenario in which these problems can be circumvented at the cost of having a low scale inflation. 
In this model, the coupling function $f$ increases during inflation starting from an initial value of 
unity and becomes very large at the end of inflation. Such an evolution of $f$ is free from the above mentioned problems. However, the coupling between the charges and EM field becomes very small at the end. To get back the standard EM theory we introduced a transition 
in the evolution of $f$ immediately after the end of inflation during which time it decreases back to unity at reheating and after that $f$ becomes constant. During this post-inflationary era both electric and magnetic energy density 
increase. By demanding that EM energy density should remain below the background energy density, we obtained a bound on reheating and inflationary scales. Our models can generate both non-helical and helical magnetic fields and satisfy known observational constraints. They predict a blue spectrum for the magnetic field energy density peaked at small length scales, typically a fraction of
the Hubble radius at reheating
\cite{sharma2017,sharmahelical}. The generated field energy density 
can also be a significant fraction of the energy density of the
Universe at those epochs.

The anisotropic stress associated with such primordial EM fields lead to a stochastic gravitational wave background. The process is to a certain degree similar
to that which obtains during a first order phase transitions in the early Universe. Prior to reheating, the electric energy density was non-zero and its amplitude is typically
greater than the magnetic energy density. Hence, prior to reheating, both electric and magnetic fields contribute to the anisotropic 
stress and result in GW production with a dominant contribution from the electric field.
Electric fields are however damped out after reheating due to the very large conductivity of the universe. Thus after reheating only the generated magnetic fields contribute to the generation of 
stochastic GW. This interplay between electric and magnetic field both contributing to 
stochastic GW leads to
a characteristic feature in the GW energy spectrum.
We calculate here the strength of the stochastic GW background generated for several 
of our inflationary magnetogenesis models. The predicted
signals are compared 
with the sensitivity of the future space based
gravitational waves detector like the Laser Interferometer Space 
Antenna (LISA) or for some reheating scales
limits obtained from Pulsar Timing Arrays (PTA). 

The paper is organised as follows. In section \ref{gwenergyspectrum} we set up the general formalism for describing the evolution 
of the stochastic GW energy spectrum in terms of the tensor perturbation of the metric.
We also introduce the different bases for representing the GW energy spectrum depending upon the nature of the source of these tensor perturbations. 
In section \ref{emtensorofthesource}, we study the inflation generated electromagnetic 
fields as the source of these perturbations and derive 
expressions for the resulting anisotropic stress needed to calculate 
the GW energy spectrum. The predicted stochastic GW spectrum due to 
non-helical electromagnetic fields is calculated in section \ref{nhemfg}. The helical case is considered in section \ref{helical}. We also compare these predictions with expected limits from 
LISA and PTA experiments. Detection of the generated GW spectrum with LISA is discussed in section \ref{detectionwlisa}. 
Some of the details of the calculations are left to several appendices.
The last section contains a discussion of our results and 
conclusions.

\section{Stochastic gravitational waves}
\label{gwenergyspectrum}

Gravitational waves may be represented by the transverse-traceless part 
of the space-time metric perturbation. These are sourced by the TT part of the energy momentum tensor. In the context of this paper, such TT component of the energy-momentum tensor is provided by the EM field. In this section we set up the general formalism to describe the evolution of 
stochastic GW energy spectrum in the expanding universe so that in subsequent sections, we can calculate the gravitational waves produced by the inflation generated EM field in our scenario. 
We consider a homogeneous, isotropic and a spatially flat background expanding universe. 
The metric for such a universe with tensor perturbation is,
\begin{align*}
ds^2=a^2(\eta)(-d \eta^2 +(\delta_{ij}+2 h_{ij})dx^idx^j) .
\end{align*}
Here $\eta$ is the conformal time, $x^i$ represents the comoving coordinates for the space dimensions, $a(\eta)$ is the scale (or expansion) 
factor and $h_{ij}$ represents the tensor perturbations of the metric, in the transverse and traceless gauge.
The energy density of the stochastic GW in terms of these tensor perturbations can be expressed as \cite{caprini2001},
\begin{align}
\rho_{GW}=\frac{1}{16 \pi G}\frac{\langle h'_{ij}h'^{ij}\rangle}{a^2} .
\end{align}
Here $h^{ij}=\delta^{im} \delta^{jn}h_{mn}$, prime ($'$) 
denotes the derivative with respect to conformal time and $\langle\rangle$ represents ensemble average. 
We define the Fourier transformation of the tensor perturbations as
\begin{equation}
h_{ij}(\vec{k},\eta)=\int {d^3 x} h_{ij}(\vec{x},\eta) e^{-i \vec{k}\cdot\vec{x}},\nonumber
\end{equation}
with the corresponding inverse transform defined as,
\begin{equation}
h_{ij}(\vec{x},\eta)=\int \frac{d^3 k}{(2 \pi)^3} h_{ij}(\vec{k},\eta) e^{i \vec{k}\cdot\vec{x}},\nonumber
\end{equation}
where we use the symbol $h_{ij}$ for both the real space and Fourier space components. The Fourier components $h_{ij}(\vec{k},\eta)$ satisfy,
$$k^i h_{ij}=0 ~~~~~~\text{and}~~~~~~~~~~~ h^i_i=0.$$
In Fourier space, $\rho_{GW}$ can be expressed as
\begin{align}\label{egwinfs}
\rho_{GW}\equiv \int d \ln k \frac{d \rho_{GW}}{d \ln k}=\frac{1}{ 16 \pi G a^2}\int \frac{d^3 k}{(2 \pi)^3}\int \frac{d^3 q}{(2 \pi)^3}\langle h'_{ij}(\vec{k},\eta) h'^{*}_{ij}(\vec{q},\eta)\rangle e^{i (\vec{k}-\vec{q})\cdot \vec{x}} .
\end{align}
We will refer to $d \rho_{GW}/d \ln k$ as the
GW energy spectrum. To estimate the energy density in GW, we need to know how $h_{ij}$ evolves with time. The evolution of $h_{ij}$ is governed by the Einstein Equation, using which, we get the following linearised equation of motion for 
$h_{ij}$ in presence of a source,
\begin{align}\label{gweom}
h''_{ij}+\frac{2 a'}{a} h'_{ij}+ k^2 h_{ij} =8 \pi G a^2 \tt_{ij} .
\end{align}
Here $a^2 \tt_{ij}$ is the transverse traceless part of the energy-momentum tensor of the source. In our case, the source is the energy momentum tensor of the electromagnetic (EM) field generated during inflation.

In our previous work \cite{sharma2017,sharmahelical}, 
we have built models of inflationary magnetogenesis which address the problems with what is referred to as the Ratra model ($f^2 F^{\mu \nu}F_{\mu \nu}$) for such generation. 
The model suggested by us is free from the strong coupling and back-reaction problem addressed in \cite{mukhanov2009} and also satisfies the constraints from the Schwinger mechanism discussed in \cite{kobayashi:2014}. In our model, inflation is followed by a matter dominated era before reheating takes place and radiation dominance starts. The coupling 
function $f$ 
grows during inflation and transits to a decaying phase after inflation until reheating. This particular evolution has been chosen to avoid the problems of strong coupling and back-reaction. In this model we show that the electromagnetic energy density is very small compared to the background energy density during inflation and it increases in the matter dominated era and becomes comparable to the background energy density at reheating. The nature of the generated electromagnetic field 
spectra at reheating is decided by the evolution of the coupling function during inflation. As the electromagnetic field gets generated, it can source the production of GW. Since the generated strength of the EM field is very small during inflation, the strength of the produced GW will also be very small. However towards the end of the pre-reheating matter dominated era the electromagnetic field although remaining below the background energy density, increases with time. This increase leads to the production of GW of a significant strength. Initially both electric and magnetic fields source the production of GW but at the end of the reheating epoch the electric field gets shorted due to the high conductivity of the constituents of the universe and after this epoch only the magnetic field contributes to the production of GW. 

Hence, for our case of interest, we need to solve the Eq.$\eqref{gweom}$ in both the matter and radiation dominated era. We consider the following evolution of scale factor during these era,
%\[
\begin{equation}
\label{aeta}
    a= 
\begin{cases}
    \frac{a_{eq}^2 H_{eq}}{4 \eta_R}(\eta+\eta_R)^2,&  \eta_e\leq\eta\leq \eta_{R}\\
    a_{eq}^2 H_{eq} \eta,              & \eta\geq \eta_{R}.
    \end{cases}
%\]
\end{equation}
Here $\eta_e$ and $\eta_R$ are the conformal time at the end of inflation and the epoch of reheating, respectively. The scale factor and Hubble parameter at the epoch of radiation-matter equaility are denoted, respectively by,
$a_{eq}$ and $H_{eq}$. The above form of the scale factor evolution in Eq.~(\ref{aeta}) ensures
the continuity of $a$ and $H$ across $\eta_R$. We need to solve for the evolution of the $h_{ij}$ sourced by the EM fields generated in the pre-reheating stage ($\eta \leq \eta_R$) and the post reheating stage ($\eta \geq \eta_R$).

\subsection{Evolution of $h_{ij}$ for $\eta \leq \eta_R$}

During the epoch $\eta_e\leq\eta\leq \eta_{R}$, we define dimensionless variable $\yy \equiv k (\eta+\eta_R)$ and 
$\Pi_{ij}\equiv[1/(\rho+p)] T_{ij}^{TT}=2(\frac{a}{a'})^2 \pi G a^2 \tt_{ij}$. In terms of these variables, Eq.~\eqref{gweom} reduces to,
\begin{align}\label{tpeom}
\frac{d^2 h_{ij}}{d \yy^2}+\frac{4}{\yy} \frac{d h_{ij}}{d \yy}+ h_{ij} = \frac{12}{\yy^2} \Pi_{ij}
\end{align}
To make the analysis simple, we choose a convenient basis to represent the tensor perturbations. The appropriate basis depends on the nature of the source of the tensor perturbation. In the next section we will see that the 
source is the anisotropic stress that
arises from either non-helical or helical EM fields. It turns out that the appropriate basis for the case of non-helical EM fields are linear polarisation basis and for helical fields are circular polarisation basis. Hence, we express the GW energy density in terms of 
amplitudes $h_{ij}$ in these different bases.

The basis set suitable to represent the linear polarisation of the gravitational waves are \cite{caprini2001},
\begin{align*}
e^{T}_{ij}&=\frac{1}{\sqrt{2}}(\hat{e}^1 \times \hat{e}^1-\hat{e}^2\times\hat{e}^2)_{ij}\\
e^{\times}_{ij}&=\frac{1}{\sqrt{2}}(\hat{e}^1 \times \hat{e}^2+\hat{e}^2\times\hat{e}^1)_{ij} .
\end{align*}
Here $(\hat{e}^1,\hat{e}^2,\hat{e}^3~\text{or}~ \hat{k})$ are a set of mutually orthonormal
basis vectors of our coordinate system and we assume that gravitational waves 
propagates in the $\hat{e}^3$ or $\hat{k}$ direction in this coordinate system. These 
basis vectors satisfy
the following properties to ensure the transverse traceless nature of the tensor perturbations,
\begin{align*}
\hat{k}^i e^{(T,\times)}_{ij}=0, \quad \delta^{ij} e^{T}_{ij}=0, \quad e^{(T,\times)}_{ij}e^{(T, \times) ij}=1.
\end{align*}
The GW tensor perturbation in terms of this basis is,
\begin{align}\label{hinlb}
h_{ij} (\vec{k},\eta)=h^{T}(\vec{k},\eta) e^{T}_{ij}+h^{\times}(\vec{k},\eta) e^{\times}_{ij}
\end{align}
Further, the suitable basis for representing the circular polarisation of the  gravitational waves are \cite{caprini2004},
\begin{align*}
e^{\pm}_{ij}=-\frac{1}{2}(\hat{e}^1\pm i\hat{e}^2)_i\times (\hat{e}^1\pm i\hat{e}^2)_j .
\end{align*}
Here $e^{\pm}_{ij}$ satisfy the following properties,
\begin{align*}
\hat{k}^i e^{\pm}_{ij}=0, \quad \delta^{ij} e^{\pm}_{ij}=0, \quad e^{\pm}_{ij}e^{\mp ij}=1.
\end{align*}
Tensor perturbations in terms of these circularly polarized basis vectors are given by,
\begin{align}\label{hincb}
h_{ij}(\vec{k},\eta)=h^{+}(\vec{k},\eta) e^{+}_{ij}+h^{-}(\vec{k},\eta) e^{-}_{ij} .
\end{align}
In terms of the appropriate basis the Eq.\eqref{tpeom} reduces to the following form,
\begin{align}
\frac{d^2 h^{\aleph}}{d \yy^2}+\frac{4}{\yy} \frac{d h^{\aa}}{d \yy}+ h^{\aa} = \frac{12}{\yy^2} \Pi^{\aa} .
\end{align}
Here ($\aa=T,\times$) or ($\aa=+,-$) for linear and circular polarisation basis respectively. The homogeneous solutions of this equation are $(j_1(\yy)/\yy)$ and $(y_1(\yy)/\yy)$ (Here $j_1(\yy)$ and $y_1(\yy)$ are first order spherical Bessel functions of first and second kind, respectively).
The complete solution of this equation is,
\begin{align} \label{hijinmd}
h^{\aa}(\vec{k},\yy) &=c_1 \frac{j_1(\yy)}{\yy}+ c_2 \frac{y_1(\yy)}{\yy}+\frac{-j_1(\yy)}{\yy} \int_{\yy_i}^\yy d \yy_1 (12 \Pi^{\aa}(\vec{k}, \yy_1)) \yy_1^2 \frac{y_1(\yy_1)}{\yy_1}+\frac{y_1(\yy)}{\yy} \int_{\yy_i}^\yy d \yy_1 (12 \Pi^{\aa}(\vec{k},\yy_1)) \yy_1^2 \frac{j_1(\yy_1)}{\yy_1}
\end{align}
Here $c_1$ and $c_2$ are constants which are determined by the matching of $h^{\aa}$ and its derivative at the epoch just before and just after the end of inflation. In our model of inflationary magnetogenesis, the spectral magnetic field energy density is proportional to the fourth power of the Hubble parameter  for a scale invariant magnetic field spectrum during inflation. Since the energy scale of inflation in our model is very low, the energy density of the magnetic field as well as the gravitational waves generated in the process  is small during inflation. The corresponding contribution to the homogeneous part of the above solution is small compared to the contribution from the source term. Therefore in the above solution, the main contribution to the GW energy density come from the terms with the source, which itself is generated during $\eta_e\leq\eta\leq \eta_{R}$.

\subsection{Evolution of $h_{ij}$ for $\eta \geq \eta_R$}
In the radiation dominated era, we define the dimensionless variable $x \equiv k \eta$ and $\Pi_{ij}$ as before. In terms of these dimensionless variables, Eq.\eqref{gweom} reduces to,
\begin{align}
\frac{d^2 h^{\aa}}{d x^2}+\frac{2}{x} \frac{d h^{\aa}}{d x}+ h^{\aa} = \frac{4}{x^2} \Pi^{\aa}.
\end{align}
The homogeneous solution of this equation are zeroth order spherical Bessel function $j_0(x)=\sin (x)/x$ and $y_0(x)=-\cos (x)/x$. 
The complete solution of this equation is,
\begin{align}\label{maiequation}
h^{\aa}(\vec{k},x)&=d_1 j_0(x)+d_2 y_0(x)-4 j_0(x) \int_{x_R}^x d x_1 \Pi^{\aa}(\vec{k}, x_1) y_0(x_1)+4 y_0(x) \int_{x_R}^x d x_1 \Pi^{\aa}(\vec{k},x_1) j_0(x_1) .
\end{align}
In the above expression, $x_R$ is the value of variable $x$ 
at reheating ($\eta=\eta_R$), while 
$d_1$ and $d_2$ are constants which are determined by matching $h_{ij}$ and its derivative at $\eta=\eta_R$.

\subsection{The GW energy spectrum}
In the next section, we will see that for statistically homogeneous
and isotropic EM fields, $\langle \Pi_{ij}(\vec{k},\eta) \Pi^{ij}(\vec{k'},\eta)$ is 
proportional to $\delta(\vec{k}-\vec{k'})$ and some function
of $k$ and $\eta$. Using this fact, the above relation 
implies that tensor perturbation $h_{ij}$ also satisfies the following property,
\begin{equation}
\langle h'_{ij}(\vec{k},\eta) h'^{ij}(\vec{k'},\eta)\rangle\propto \delta(\vec{k}-\vec{k'})\times F(k,\eta)
\end{equation}
for some function $F(k,\eta)$.
From this property and using the expansions given
in Eq.\eqref{hinlb} and in Eq.\eqref{hincb}, 
we express $\langle h'_{ij}(\vec{k},\eta)h'^{ij*}(\vec{k'},\eta)\rangle$ as,
\begin{equation}\label{hijindb}
\langle h'_{ij}(\vec{k},\eta)h'^{ij*}(\vec{k'},\eta)\rangle= (2\pi)^3\delta(\vec{k}-\vec{k'})\left(\Big|\frac{dh^{T}(k,\eta)}{d \eta}\Big|^2+ \Big|\frac{d h^{\times}(k,\eta)}{d \eta}\Big|^2\right)=(2\pi)^3\delta(\vec{k}-\vec{k'})\left(\Big|\frac{dh^{+}(k,\eta)}{d \eta}\Big|^2+ \Big|\frac{d h^{-}(k,\eta)}{d \eta}\Big|^2\right).
\end{equation}

After substituting Eq.\eqref{hijindb} in Eq.\eqref{egwinfs}, we get
\begin{align}\label{GWpectrum}
\frac{d\rho_{GW}}{d \ln k}=\frac{k^3}{4(2 \pi)^3 G a^2}\sum_{\aa}\left(\Big|\frac{dh^{\aa}(k,\eta)}{d \eta}\Big|^2\right)
\end{align}

Further, after normalising the gravitational energy density with background energy density at present ($\rho_{c_0}$), we get
\begin{align}\label{gwenergydensity}
\frac{d\Omega_{GW}}{d \ln k}\Bigg|_0=\frac{d\Omega_{GW}}{d \ln k}\Bigg|_{\eta} a^4(\eta)=\frac{k^3 a^2}{4(2 \pi)^3 G \rho_{c_0}}\sum_{\aa}\left(\Big|\frac{dh^{\aa}(k,\eta)}{d \eta}\Big|^2\right),
\end{align}
where, in the above expression we have, as before, defined as,
$d\Omega_{GW}/d \ln k=(1/\rho_{c_0}) d \rho_{GW}/d \ln k$.

\section{Energy momentum tensor of the source}\label{emtensorofthesource}
To calculate the GW energy spectrum we need to calculate the anisotropic stress tensor of the source. The energy momentum tensor of the electromagnetic field is given by,
\begin{align*}
T_{\mu \nu}=\frac{1}{4 \pi}\left(g^{\alpha \beta}F_{\mu \alpha} F_{\nu \beta} -\frac{g_{\mu \nu}}{4}F^{\alpha \beta} F_{\alpha \beta}\right).
\end{align*}
Anisotropic stress tensor is given by the transverse traceless projection of the spatial part of the energy momentum tensor. Spatial part of the energy momentum tensor is,
\begin{align}\label{tij}
T_{ij}(\vec{x},\eta)&=\frac{1}{4 \pi}\big(B_i(\vec{x},\eta) B_j(\vec{x},\eta)+E_i(\vec{x},\eta) E_j(\vec{x},\eta) -\frac{1}{2} g_{ij} B^m B_m -\frac{1}{2} g_{ij} E^m E_m\big)
\end{align}
where $$E_i=\frac{1}{a} F_{i 0} =-\frac{1}{a} A'_i~~~~~\text{and}~~~~B_{i}=\frac{1}{2a}\epsilon^*_{ijk}\delta^{jl}\delta^{km} F_{l m}=\frac{1}{a}\epsilon^*_{ijk}\delta^{jl}\delta^{km} \partial_l A_m$$ are the covariant components of the electric and magnetic field with respect to the comoving observer with four velocity $u^{\mu}\equiv(1/a,0,0,0)$ \cite{kandu2010}. Here $A_i$ the is spatial part of the EM $4$-potential and $\epsilon^*_{ijk}$ is 3-d fully antisymmetric symbol with $\epsilon^*_{123}=1$. After taking the Fourier transformation of the Eq.\eqref{tij}, we get
\begin{align}
T_{ij}(\vec{k},\eta)&=\frac{1}{4 \pi}\Big(\int \frac{d^3 q}{(2 \pi)^3}  B_{i}(\vec{q},\eta) B^*_{j}(\vec{q}-\vec{k},\eta)+\int \frac{d^3 q}{(2 \pi)^3}  E_{i}(\vec{q},\eta) E^*_{j}(\vec{q}-\vec{k},\eta)  \Big)-\frac{1}{8\pi}\Big(g_{ij}\int \frac{d^3 q}{(2 \pi)^3}(B_{a}(\vec{q},\eta) B^{*a}(\vec{q}-\vec{k},\eta)\nonumber\\
&+E_{a}(\vec{q},\eta) E^{*a}(\vec{q}-\vec{k},\eta))\Big) \label{emt}
\end{align}
The transverse traceless part of the $T_{ij}$ is given by,
\begin{align*}
a^2 \tt_{ij}(\vec{k},\eta)&=\frac{1}{4 \pi}\Bigg(\int \frac{d^3 q}{(2 \pi)^3} P_{ij}^{m n} \Big(B_{m}(\vec{q},\eta) B^*_{n}(\vec{q}-\vec{k},\eta)+E_{m}(\vec{q},\eta) E^*_{n}(\vec{q}-\vec{k},\eta)\Big)  \Bigg)
\end{align*}

Here $P_{ij}^{m n}=P_i^m P_j^n-1/2 P_{ij} P^{m n}$. In the above we do not take the contribution from the term in the second bracket of Eq.(\ref{emt}) because that term does not contribute to the transverse traceless part. We are interested in the evolution of $\langle T^{TT}_{ij}(\vec{k},\eta)\tt^{*ij}(\vec{k'},\eta')\rangle$ to calculate $\rho_{GW}$. This is given by
\begin{align*}
\tt_{ij}(\vec{k},\eta)\tt^{*kl}(\vec{k'},\eta')=&\left(\frac{1}{4 \pi}\right)^2 \left(\int \frac{d^3 q}{(2 \pi)^3 a^2(\eta)} P_{ij}^{m n} \left(B_{m}(\vec{q},\eta) B^*_{n}(\vec{q}-\vec{k},\eta)+E_{m}(\vec{q},\eta) E^*_{n}(\vec{q}-\vec{k},\eta)\right)   \right)\nonumber\\
&\left(\int \frac{d^3 q' a^4(\eta')}{(2 \pi)^3 a^2(\eta')} P^{kl}_{a b}\left(B^{a}(\vec{q'},\eta') B^{*b}(\vec{q'}-\vec{k'},\eta')+E^{a}(\vec{q'},\eta') E^{*b}(\vec{q'}-\vec{k'},\eta')\right)  \right)
\end{align*}
We will now proceed by expressing $\vec{B},~\vec{E}$ and $\rho$ in terms of their corresponding comoving values. Taking expectation value of the product of $\tt_{ij}(\vec{k},\eta)$'s, we get
\begin{align}
\langle\tt_{ij}(\vec{k},\eta)\tt^{*kl}(\vec{k'},\eta')\rangle=&\left(\frac{1}{4 \pi a^4(\eta)}\right) \left(\frac{1}{4 \pi a^4(\eta')}\right)\Bigg(\int \frac{d^3 q}{(2 \pi)^3}\int \frac{d^3 q'}{(2 \pi)^3}P_{ij}^{m n}   P^{kl}_{a b}\Big(\langle\tilde{B}_{m}(\vec{q},\eta) \tilde{B}^*_{n}(\vec{q}-\vec{k},\eta) \nonumber\\
& \tilde{B}^{*a}(\vec{q'},\eta') \tilde{B}^{b}(\vec{q'}-\vec{k'},\eta')  \rangle+\langle \tilde{E}_{m}(\vec{q},\eta) \tilde{E}^*_{n}(\vec{q}-\vec{k},\eta)\tilde{E}^{*a}(\vec{q'},\eta') \tilde{E}^{b}(\vec{q'}-\vec{k'},\eta')\rangle\Big) \Bigg)\label{coreqn}.
\end{align} 
In the above expression, tilde over the quantities represents their comoving values ($ \tilde{B}_{b}(\vec{q},\eta)\equiv B_b(\vec{q},\eta)/a(\eta)$ and $ \tilde{B}^{b}(\vec{q},\eta)\equiv B^b(\vec{q},\eta)/a^3(\eta)$). We have neglected the contribution of the cross terms of electric and magnetic field because those terms are always subdominant for our case of interest.
%%%%%%%%%%%%%%%%%%%%%%%%%%%%%%%%%%%%%%%%%%%%%%%%%%%%%%%%%%%%%%%%%%%\\
%%%%%%%%%%%%%%%%%%%%%%%%%%%%%%%%%%%%%%%%%%%%%%%%%%%%%%%%%%%%%%%%%%%\\
%%%%%%%%%%%%%%%%%%%%%%%%%%%%%%%%%%%%%%%%%%%%%%%%%%%%%%%%%%%%%%%%%%%
From Eq. \eqref{coreqn} it is clear that calculations of $\langle\tt_{ij}(\vec{k},\eta)\tt^{*kl}(\vec{k'},\eta')\rangle$ involves $(\langle\tilde{B}_{m}(\vec{q},\eta) \tilde{B}^*_{n}(\vec{q}-\vec{k},\eta)  \tilde{B}^{*a}(\vec{r},\eta') \tilde{B}^{b}(\vec{r}-\vec{k'},\eta')  \rangle$. Since the nature of the magnetic field generated in our model is gaussian, we can express these four point correlation functions in terms of the two point correlation functions.
\begin{align}
\langle\tilde{B}_{m}(\vec{q},\eta) \tilde{B}^*_{n}(\vec{q}-\vec{k},\eta)  \tilde{B}^{*a}(\vec{r},\eta') \tilde{B}^{b}(\vec{r}-\vec{k'},\eta')  \rangle=&\langle\tilde{B}_{m}(\vec{q},\eta) \tilde{B}^*_{n}(\vec{q}-\vec{k},\eta)\rangle \langle  \tilde{B}^{*a}(\vec{r},\eta') \tilde{B}^{b}(\vec{r}-\vec{k'},\eta')  \rangle\nonumber\\
&+\langle\tilde{B}_{m}(\vec{q},\eta) \tilde{B}^{*a}(\vec{r},\eta')\rangle \langle \tilde{B}^*_{n}(\vec{q}-\vec{k},\eta)  \tilde{B}^{b}(\vec{r}-\vec{k'},\eta')  \rangle\nonumber\\
&+  \langle\tilde{B}_{m}(\vec{q},\eta)\tilde{B}^{b}(\vec{r}-\vec{k'},\eta')\rangle \langle \tilde{B}^*_{n}(\vec{q}-\vec{k},\eta)  \tilde{B}^{*a}(\vec{r},\eta')   \rangle \label{fourpoint}
\end{align}
In Eq.(\ref{fourpoint}), we require unequal time correlation of the magnetic fields. For this we followed the analysis in \cite{caprini2007,sigl2018} and represent the unequal time correlation of the source in terms of the product of the 
equal time correlation, $\langle \tilde{B}_{i}(\vec{k},\eta) \tilde{B}_{j}(\vec{k'},\eta)\rangle$
and a two time
correlation function $C_B(k,\eta,\eta')$, which depends on these different times as follows,
\begin{equation} \label{dtcf}
\langle \tilde{B}_{i}(\vec{k},\eta) \tilde{B}_{j}(\vec{k'},\eta')\rangle=\langle \tilde{B}_{i}(\vec{k},\eta) \tilde{B}_{j}(\vec{k'},\eta)\rangle C_B(k,\eta,\eta').
\end{equation}
It is evident from the above relation that for equal time correlation $C_B(k,\eta,\eta)=1$. To proceed further, we need to know the equal time correlation function of the electric and magnetic field. We divide the further study in two parts depending upon the nature of the generated EM field; non-helical and helical nature.

\subsection{Non helical EM fields}

For non helical magnetic  and electric fields, we represent the two point correlation function in terms of the power spectrum as follows \cite{caprini2009},
\begin{align}
\langle \tilde{B}_{i}(k,\eta)\tilde{B}^*_{j}(k',\eta) \rangle&= (2 \pi)^3 \delta(\vec{k}-\vec{k'})(\delta_{ij}-\hat{k}_i \hat{k}_j)P_{SB}(k,\eta)\nonumber\\
\langle \tilde{E}_{i}(k,\eta) \tilde{E}^*_{j}(k',\eta) \rangle&= (2 \pi)^3 \delta(\vec{k}-\vec{k'})(\delta_{ij}-\hat{k}_i \hat{k}_j)P_{SE}(k,\eta) \label{tpce}
\end{align}
In the above expression, we have assumed that the distribution of the generated electric and magnetic field is homogeneous and isotropic. The delta function, $\delta(\vec{k}-\vec{k'})$ in the above expression and the dependence of power spectrum $P_{SB}$ only on the the magnitude of the $\vec{k}$ arise because of the this homogeneous and isotropic nature of the electromagnetic field distribution. The projection tensor ($\delta_{ij}-\hat{k}_i \hat{k}_j$) in the above expression ensures the divergence less nature of the magnetic field. 
We also have this projection tensor in the electric field correlation 
function as during the EM field generation, charge particles density is negligible. Hence, the electric field can be assumed to have zero divergence.

Equations \eqref{coreqn},\eqref{fourpoint},\eqref{dtcf} and \eqref{tpce} imply
\begin{align}\label{piijnh}
\langle\tt_{ij}(\vec{k},\eta)\tt^{*ij}(\vec{k'},\eta')\rangle=&\frac{1}{a^4(\eta) a^4(\eta')}\left(f_B(k,\eta,\eta')+f_E(k,\eta,\eta')\right)(2 \pi)^3\delta(\vec{k}-\vec{k'}).
\end{align}
Here,
\begin{align}
f_{B,E}(k,\eta,\eta')=& \frac{1}{4(2\pi)^5}\int d^3 q  \Big[ P_{SB,SE}(q,\eta)P_{SB,SE}({|\vec{k}-\vec{q}|},\eta) (1+\gamma^2+\beta^2+\gamma^2 \beta^2)\Big]\nonumber\\
&C_{B,E}(q,\eta,\eta') C_{B,E}(|\vec{k}-\vec{q}|,\eta,\eta'). \label{fbfe}
\end{align} 
In the above expression $\gamma=\hat{k}\cdot \hat{q}$ and $\beta=\hat{k}\cdot \widehat{k-q}$. The detailed derivation
of the above expression is given in Appendix \eqref{appendixa}.

To get the individual mode contribution, we express $\tt_{ij}(\vec{k},\eta)$ in terms of the linear polarisation basis.
\begin{align*}
\tt_{ij}(\vec{k},\eta)=\tt^T(\vec{k},\eta) e_{ij}^T +\tt^\times(\vec{k},\eta) e_{ij}^\times
\end{align*}
Using this we get,
\begin{align}\label{iminnh}
\langle\tt_{ij}(\vec{k},\eta)\tt^{*ij}(\vec{k'},\eta')\rangle=&(|\tt^T|^2(k,\eta,\eta')+|\tt^\times|^2(k,\eta,\eta'))(2\pi)^3\delta(\vec{k}-\vec{k'}).
\end{align}
In this case, the source is such that the contribution to both the modes 
($T~ \text{and}~ \times$) are equal. 
From Eq.~(\ref{piijnh}) and Eq.~(\ref{iminnh}), we get,
\begin{align*}
|\tt^T|^2(k,\eta,\eta')=|\tt^\times|^2(k,\eta,\eta')&=\frac{1}{2}\frac{1}{a^4(\eta) a^4(\eta')}\Big(f_B(k,\eta,\eta')+f_E(k,\eta,\eta')\Big).
\end{align*} 
%%%%%%%%%%%%%%%%%%%%%%%%%%%%%%%%%%%%%%%%%%%%%%%%%%%%%%%%%%%%%%%%%%%\\
%%%%%%%%%%%%%%%%%%%%%%%%%%%%%%%%%%%%%%%%%%%%%%%%%%%%%%%%%%%%%%%%%%%\\
%%%%%%%%%%%%%%%%%%%%%%%%%%%%%%%%%%%%%%%%%%%%%%%%%%%%%%%%%%%%%%%%%%%\\
%%%%%%%%%%%%%%%%%%%%%%%%%%%%%%%%%%%%%%%%%%%%%%%%%%%%%%%%%%%%%%%%%%%
\subsection{Helical EM fields}
We follow a similar procedure for the case of helical field. The only difference that will arise is that there is an additional antisymmetric contribution to the 2-point correlation function. For helical EM field, we have \citet{caprini2004},
\begin{align}
\langle \tilde{B}_{i}(k,\eta)\tilde{B}^*_{j}(k',\eta) \rangle&= (2 \pi)^3 \delta(\vec{k}-\vec{k'})\Big((\delta_{ij}-\hat{k}_i \hat{k}_j)P_{SB}(k,\eta)+i \epsilon_{ijm}\hat{k}_m P_{AB}(k,\eta) \Big)\nonumber\\
\langle \tilde{E}_{i}(k,\eta)\tilde{E}^*_{j}(k',\eta) \rangle&= (2 \pi)^3 \delta(\vec{k}-\vec{k'})\Big((\delta_{ij}-\hat{k}_i \hat{k}_j)P_{SE}(k,\eta)+i \epsilon_{ijm}\hat{k}_m P_{AE}(k,\eta) \Big)\label{tpceh}
\end{align}
The term containing $P_{AB}, P_{AE}$ are the antisymmetric parts of two point correlation tensor.\\
Equations \eqref{tpceh},\eqref{fourpoint} and \eqref{coreqn} imply
\begin{align}\label{piijh}
\langle\tt_{ij}(\vec{k},\eta)\tt^{*ij}(\vec{k'},\eta')\rangle&=\frac{1}{a^4(\eta) a^4(\eta')}\left(g_B(k,\eta,\eta')+g_E(k,\eta,\eta')\right)(2 \pi)^3 \delta(\vec{k}-\vec{k'}),
\end{align}
where,
\begin{align}
g_{B,E}(k,\eta,\eta')=& \frac{1}{4(2\pi)^5}\int d^3 q  \Bigg[ P_{SB,SE}(q,\eta)P_{SB,SE}({|\vec{k}-\vec{q}|},\eta) (1+\gamma^2+\beta^2+\gamma^2 \beta^2)\nonumber\\
&+4 \gamma \beta P_{AB,AE}(q,\eta)P_{AB,AE}({|\vec{k}-\vec{q}|},\eta)\Bigg]C_{B,E}(q,\eta,\eta') C_{B,E}(|\vec{k}-\vec{q}|,\eta,\eta'). \label{gbge}
\end{align} 
Here $\gamma=\hat{k}\cdot \hat{q}$ and $\beta=\hat{k}\cdot \widehat{k-q} $. The detailed derivation
of the above expression is given in Appendix \eqref{appendixb}.

To write down the individual mode contribution for this case, we express $\Pi_{ij}(\vec{k})$ in terms of the circular polarisation basis.
\begin{align*}
\tt_{ij}(\vec{k},\eta)&=\tt^+(k,\eta) e_{ij}^+ +\tt^-(k,\eta) e_{ij}^-\\
\langle\tt_{ij}(\vec{k},\eta)\tt^{*ij}(\vec{k'},\eta')\rangle&=(|\tt^+|^2(k,\eta,\eta')+|\tt^-|^2(k,\eta,\eta'))(2\pi)^3\delta(\vec{k}-\vec{k'})
\end{align*}
In this case, the individual mode ($+ , - $) not only involves the terms arising from the terms containing $P_S P_S$ and $P_A P_A$ but also the cross term containing $P_S P_A$. These terms contribute to individual mode as follows,
\begin{align*}
|\tt^-|^2(k,\eta,\eta')&=\frac{1}{2}\frac{1}{a^4(\eta) a^4(\eta')}(g_B(k,\eta,\eta')+g_E(k,\eta,\eta')+h_B(k,\eta,\eta')+h_E(k,\eta,\eta'))\nonumber\\
\text{and}~~~~~|\tt^+|^2(k,\eta,\eta')&=\frac{1}{2}\frac{1}{a^4(\eta) a^4(\eta')}(g_B(k,\eta,\eta')+g_E(k,\eta,\eta')-h_B(k,\eta,\eta')-h_E(k,\eta,\eta')).
\end{align*}
Here
\begin{align*}
h_{B,E}(k,\eta,\eta')=&\frac{1}{4(2\pi)^5}\int d^3 q  \Big[  P_{SB,SE}(q,\eta) P_{AB,AE}({|\vec{k}-\vec{q}|},\eta)4 (1+\gamma^2) \beta\Big] C_{B,E}(q,\eta,\eta')C_{B,E}(|\vec{k}-\vec{q}|,\eta,\eta') 
\end{align*}

%%%%%%%%%%%%%%%%%%%%%%%%%%%%%%%%%%%%%%%%%%%%%%%%%%%%%%%%%%%%%%%%%%%%%%%%%%%%%%
%%%%%%%%%%%%%%%%%%%%%%%%%%%%%%%%%%%%%%%%%%%%%%%%%%%%%%%%%%%%%%%%%%%%%%%%%%%%%%\\
%%%%%%%%%%%%%%%%%%%%%%%%%%%%%%%%%%%%%%%%%%%%%%%%%%%%%%%%%%%%%%%%%%%%%%%%%%%%%%\\
%%%%%%%%%%%%%%%%%%%%%%%%%%%%%%%%%%%%%%%%%%%%%%%%%%%%%%%%%%%%%%%%%%%%%%%%%%%%%%\\
%%%%%%%%%%%%%%%%%%%%%%%%%%%%%%%%%%%%%%%%%%%%%%%%%%%%%%%%%%%%%%%%%%%%%%%%%%%%%%\\
%%%%%%%%%%%%%%%%%%%%%%%%%%%%%%%%%%%%%%%%%%%%%%%%%%%%%%%%%%%%%%%%%%%%%%%%%%%%%%

\section{Predicted Gravitational Wave Spectrum}
\label{generatedgw}

In our model, to address the strong coupling and back-reaction problems of inflationary magnetogenesis, we have taken a particular evolution of the coupling function, $f$ which evolves with time both during as well as in the era after inflation till reheating. This function increases during inflation and transits to a decaying phase post inflation. We have assumed that the era between the end of inflation and the beginning of reheating is matter dominated. After this matter dominated era, reheating takes place and standard radiation dominance starts. During inflation the magnetic field spectrum is scale invariant but the strength is very low compared to the background energy density because of the low scale of inflation. In the post inflation era when coupling function, $f$ decreases, the scale invariant contribution to the magnetic spectrum decreases but contribution from the next order gets amplified on the superhorizon scales. This post inflationary era ends when the EM energy density is $\epsilon$ times the background energy density and after this reheating takes place and EM energy density evolves like radiation. The magnetic field spectrum generated in our model 
is a blue spectrum, $d \tilde{\rho}_{B}(k,\eta)/d \ln k
\propto k^4$, where $\tilde{\rho}_B$ is the comoving magnetic
energy density.

The main contribution to the GW energy spectrum takes place during the end phase of the post inflationary matter dominated era. During this era both electric and magnetic fields contribute to
the production of GW. However after reheating, electric fields get shorted out because of the large conductivity of the universe and only magnetic field contributes to the production of GW.  We have considered scenarios of magnetogenesis where non helical fields are generated \cite{sharma2017} as well as a scenario \cite{sharmahelical} where the EM field generated is almost fully helical. In subsequent sections, we therefore consider
GW energy spectrum generated due to both
non-helical and helical EM fields.

Using Eq.\eqref{gwenergydensity}, GW energy spectrum can be expressed as,
\begin{align}
\frac{d\Omega_{GW}}{d \ln k}\Bigg|_0&=\frac{k^3 a^2}{4(2 \pi)^3 G \rho_c} \sum_\aa\left(\Big|\frac{dh^{\aa}(\eta)}{d \eta}\Big|^2\right)=\frac{\Omega_R k^3 x^2}{12 \pi^2} \sum_\aa\left(\Big|\frac{dh^{\aa}(x)}{d x}\Big|^2\right)\label{omegainhp},
\end{align}
where we can calculate 
$|dh^{\aa}(x)/dx|^2$ using Eq.\eqref{maiequation}. In the limit $x>>1$, we get
\begin{align}
\Big|\frac{dh^{\aa}(k,x)}{d x}\Big|^2=&\frac{1}{2 x^2}(|d_1|^2+|d_2|^2)+\frac{8}{x^2}\int_{x_R}^{x_{\nu d}} \int_{x_R}^{x_{\nu d}} \frac{d x_1 d x_2}{x_1 x_2} \cos(x_2-x_1) |\Pi^\aa|^2(k,x_1,x_2)\label{hijexp}
\end{align}
The calculation of the above expression is given in Appendix \ref{hcalculation}. The expression for $|\Pi^\aa|^2$ is as given in Eq.~\eqref{expforpi}. In the above expression, for the second term the limits of the integration are from the epoch of reheating to the neutrino decoupling epoch ($x_{\nu d}= k \eta_{\nu d}$) and only magnetic field contributes for this case as electric field gets shorted out by the large conductivity of the universe after reheating. After neutrino decoupling epoch, anisotropic stress of the magnetic field is balanced by the anisotropic stress of the neutrinos \cite{lewis2004} and there is no further production of GW take place. The expressions for $|d_1|^2$ and $|d_2|^2$ contains the $| \Pi^{\aa}|^2(k,\eta,\eta')$ and also the different time correlation function $C_{B,E}(k,\eta,\eta')$. To evaluate $| \Pi^{\aa}|^2(k,\eta,\eta')$, we need to know two point correlation of electric and magnetic fields which takes different forms for non helical and helical EM field as discussed in the section~\ref{emtensorofthesource}. 
Therefore, we perform further analysis in two parts depending upon 
the non-helical and helical nature of the EM field.

\subsection{Gravitational waves energy spectrum for non helical magnetic field}\label{nhemfg}

To evaluate the GW energy spectrum, we need to evaluate $|\Pi^\aa|^2(k,\eta,\eta')$. Using $\Pi_{ij}=1/(\rho+p)T_{ij}^{TT}$, we express $|\Pi^\aa|^2$ in terms of $|\tt^{\aa}|^2$ which we have calculated in the section~\ref{emtensorofthesource} and we get,
\begin{equation}\label{expforpi}
|\Pi^\aa|^2 (k,\eta,\eta')=\frac{1}{(\rho+p)(\eta)}\frac{1}{(\rho+p)(\eta')}|\tt^{\aa}|^2(k,\eta,\eta')
\end{equation}
In the matter dominated era before reheating, ($\rho+p) \propto a^{-3}$, whereas, in the radiation dominated era post reheating, we have, ($\rho+p) \propto a^{-4}$. Using this, the above relation reduces to the following expression for non-helical field,
\begin{equation} \label{piinnonhelical}
|\Pi^\aa|^2 (k,\eta,\eta') = \left\{
        \begin{array}{ll}
           \frac{a_R}{a(\eta)}\frac{a_R}{a(\eta')}\left(\frac{1}{\tilde{\rho}+\tilde{p}}\right)^2\frac{1}{2}\Big(f_B(k,\eta,\eta')+f_E(k,\eta,\eta')\Big),&  \eta, \eta'\leq\eta_R\\
          \left(\frac{1}{\tilde{\rho}+\tilde{p}}\right)^2 \frac{f_B(k,\eta,\eta')}{2},&  \eta, \eta'\geq\eta_R
        \end{array}
    \right.
\end{equation}
In the above expression tilde over quantities represents their comoving values.
As is evident from Eq.\eqref{fbfe}, to calculate $f_B$ and $f_E$, we need to know the electric and magnetic field power spectrum in the matter dominated era after inflation and their evolution after reheating. These power spectra can be expressed in the form of spectral energy density of magnetic and electric fields as follows,
\begin{align}\label{defpowerspec}
P_{SB,SE}(k,\eta)=\frac{(2\pi)^3}{k^3}\frac{d \tilde{\rho}_{B,E}(k,\eta)}{d \ln k}
\end{align}
During the matter dominated era the electric and magnetic spectral energy density increase at a rate decided by how the coupling function decreases. In our model discussed in \citet{sharma2017}, the coupling function $f\propto a^{-\beta}$ ($\beta=2N/N_r$ where $N$ and $N_r$ are the number of e-folds during inflation and after the end of inflation to reheating, respectively) and the comoving spectral electric and magnetic field energy density evolve as,
\begin{equation}\label{ms}
\frac{d \tilde{\rho}_B(k, \eta)}{d \ln k} = \left\{
        \begin{array}{ll}
           D_1 \left(\frac{k}{\kp(\eta)}\right)^4\left(\frac{\eta+\eta_R}{2\eta_R}\right)^{8 \beta+2},&  k\leq \kp(\eta),\quad \eta\leq\eta_R\\
           D_1 \left(\frac{k}{\kp(\eta)}\right)^4\left(\frac{\eta_k+\eta_R}{2\eta_R}\right)^{8 \beta+2},&  k\geq \kp(\eta),\quad \eta\leq\eta_R
        \end{array}
    \right.
\end{equation}
\begin{equation}\label{es}
    \frac{d \tilde{\rho}_E(k,\eta)}{d \ln k}= \left\{
    \begin{array}{ll}
D_2 \left(\frac{k}{\kp(\eta)}\right)^2 \left(\frac{\eta+\eta_R}{2\eta_R}\right)^{8 \beta},&  k\leq \kp(\eta),\quad \eta\leq\eta_R\\
 D_2 \left(\frac{k}{\kp(\eta)}\right)^2\left(\frac{\eta_k+\eta_R}{2\eta_R}\right)^{8 \beta},&  k\geq \kp(\eta),\quad \eta\leq\eta_R
\end{array}
\right.
\end{equation}
Here $\kp(\eta)$ is the mode where electric and magnetic spectral energy density peak. For the model discussed in \citep{sharma2017}, $k_p(\eta)=\beta k_H(\eta)$ \footnote{In Ref. \citep{sharmajuly2017, sharmahelical}, we have taken $\beta\approx1$ for the wavenumber where the magnetic and electric spectrum peak. However, for the calculation of the GW spectrum , we have taken the actual value of $\beta$.} where $k_H(\eta)$ is the mode corresponding to the horizon size at conformal time $\eta$. $\eta_R$ is the epoch of reheating, and, $D_1$ and $D_2$ are, respectively, the amplitudes of spectral magnetic and electric energy densities at $k_0=\kp(\eta_R)$ which is the comoving horizon scale at the epoch of reheating, denoted by the conformal time, $\eta=\eta_R$. Values of $D_1$ and $D_2$ depend on the fraction of electromagnetic energy density to background energy density at reheating and $D_2$ is 4 times the value of $D_1$ in our model of inflationary magnetogenesis. The above expression for the case $k\leq \kp$ has been derived in the Ref.\cite{sharma2017}. For the modes which enter during the matter dominated era ($k\geq k_0$), we approximate their spectral energy density by the value at $\eta=\eta_k$ when the mode enters the horizon. The contribution of these modes will not make much difference to the GW spectrum. 

For this case, we know the exact time evolution of the EM field during the matter dominance era (see Ref. \cite{sharma2017} for details). Thus we express the two point correlation function at different times in terms of the power spectrum of the electromagnetic fields with the help of the following correlation function,
\begin{equation*}
C_B(k,\eta, \eta') = \left(\frac{\eta'+\eta_R}{\eta+\eta_R}\right)^{4 \beta+1} ~~~~~~~~~ \text{and}~~~~~~~
C_E(k,\eta, \eta') = 
     \left(\frac{\eta'+\eta_R}{\eta+\eta_R}\right)^{4 \beta} \quad \text{for}~~~~~~~~~ \eta,\eta'<\eta_R
\end{equation*}
After reheating, electric field does not contributes to GW spectrum as it gets shorted out due to the large conductivity of the universe. The spectrum of the magnetic field energy density at reheating is a 
blue spectrum which peaks at $k=k_0$ at reheating. 
After reheating the universe enters to the radiation dominated era from the matter dominated era. Larger and larger scale
superhorizon modes begin to enter the horizon. Non-linear processing of the magnetic field energy density becomes important when the Alfven crossing time becomes equal to the Hubble time i.e. $k V_A (k) = a H$ \cite{kandu2016,jedamzik}. Here $V_A=\sqrt{(d \tilde{\rho}_B(k)/d \ln k)/(\tilde{\rho}+\tilde{p})}$ is the Alfven velocity for the mode $k$.
For simplicity, we assume that non-linear processing starts just after reheating. After the onset of nonlinear evolution of the magnetic field, the detailed 
analysis of their
evolution requires numerical simulation \cite{axel, axel2017, zrake}. Further,
the calculation of the GW spectrum also requires numerical simulation which has been recently done in Ref.\cite{axelgw} for the magnetohydrodynamic turbulance in the early universe. Here we use the analytical results for the evolution of magnetic field energy density discussed in \cite{jedamzik, kandu2016},
\begin{equation}\label{rhobinnh}
   \frac{d \tilde{\rho}_B(k,\eta)}{d \ln k}= \left\{
    \begin{array}{ll}
 D_{1} \left(\frac{k}{k_{NL}(\eta)}\right)^4 \left(\frac{\eta}{\eta_R}\right)^{-\frac{4}{3}},  k\leq k_{NL}(\eta)\\
    D_{1} \left(\frac{k}{k_{NL}(\eta)}\right)^{-\frac{2}{3}} \left(\frac{\eta}{\eta_R}\right)^{-\frac{4}{3}}, 
k_{\nu}\geq k\geq k_{NL}(\eta)
\end{array}
\right.
\end{equation}
where,
\begin{equation}
k_{NL}(\eta) = k_0 \left(\frac{\eta}{\eta_R}\right)^{-\frac{1}{3}}.
\end{equation}
Here $k_{\nu}$ 
is the wave number above which viscosity dominate and the spectrum becomes exponentially damped. We will refer to the branch of the magnetic spectrum, which develops due to
the MHD turbulent cascade of energy to smaller and smaller scales, and 
with $d \tilde{\rho}_B/{d \ln k} \propto k^{-2/3}$ as the `Kolmogorov' branch.
For the estimation of GW energy 
spectrum, we also need to know the unequal time correlation function of the magnetic field energy densities. 
To find the unequal time correlation requires numerical simulation. 
It has been approximated in Ref. \cite{sigl2018} by the following expression, which we adopt,
\begin{equation}
\label{cornonlin}
  C_B(k,\eta_1,\eta_2)= \left\{
    \begin{array}{ll}
\exp\left[\frac{-(\eta_1-\eta_2)^2}{2 \tau_{E}^2(k,\eta_{max})}\right],&  k_{\nu}\geq k\geq k_{NL}(\eta_{max})\\
    1,              &  k\leq k_{NL}(\eta_{max}) .
\end{array}
\right.
\end{equation}

Here $\eta_{max}=Max[\eta_1,\eta_2]$ and 
\begin{equation}
\tau_{E}(k,\eta)=\frac{1}{k \sqrt{\frac{1}{(\tilde{\rho}+\tilde{p})}\left\langle\frac{ d \tilde{\rho}_B(k)}{d \ln k}\right\rangle}}\label{eddytime}
\end{equation}
is the
eddy turnover time for the mode $k$, assuming that these eddies have developed a velocity comparable to the Alfv\'en velocity. 
In the above expression
$$\left\langle\frac{ d \tilde{\rho}_B(k)}{d \ln k}\right\rangle=\int_{0}^{\infty} d \ln k \frac{ d \tilde{\rho}_B(k)}{d \ln k}. $$

After substituting the expression of $ d \tilde{\rho}_B(k)/d \ln k$ from Eq.\eqref{rhobinnh} in Eq.\eqref{eddytime}, we get
\begin{align}
\tau_{E}^2(k,\eta)&\approx \frac{2}{3 k^2} \frac{\tilde{\rho}+\tilde{p}}{D_1}\left(\frac{\eta}{\eta_R}\right)^{4/3} .
\end{align}
Using the information of EM energy densities and correlation function in Eq.~(\ref{cornonlin}) applicable
to this case, we evaluate the expression given in Eq.\eqref{hijexp}. To calculate GW energy spectra, we need to solve the integrals in Eq.$\eqref{hijexp}$ and substitute its value in Eq.\eqref{omegainhp}. The exact estimation of these integrals cannot be done analytically and we will present
numerical calculations below. However, one can get analytical estimate for the modes with $k<k_H$, which we can compare with corresponding numerical results. Before this, and to compare partially
with
the numerical results, we now consider an analytical estimate.
\subsubsection{Analytical Estimates}
\begin{figure*}
\includegraphics[scale=0.35]{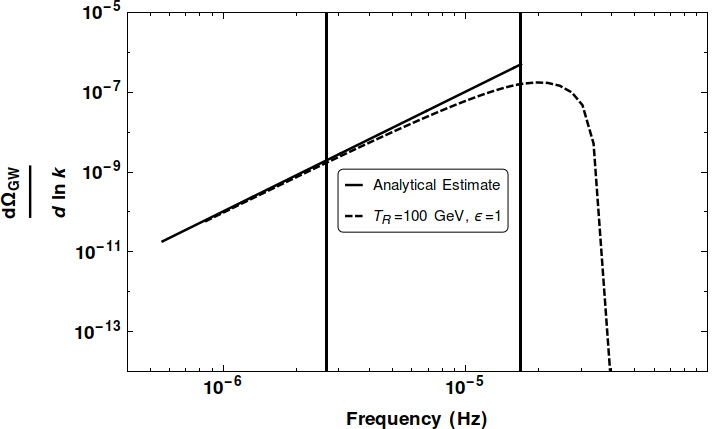}
\includegraphics[scale=0.35]{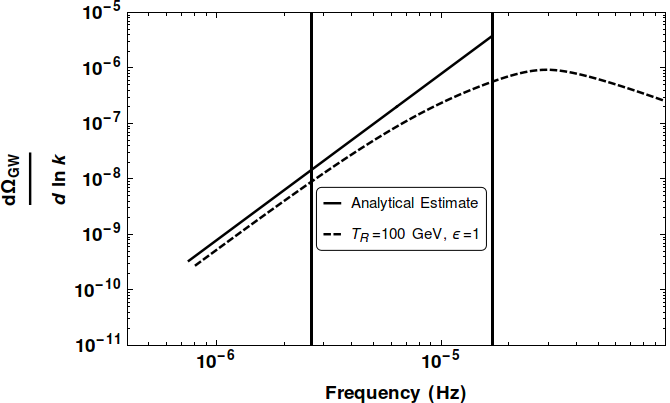}
  \caption{In this figure we plot the GW energy spectrum obtained from the numerical calculation along with the analytical estimate given in Eq.\eqref{analytical}. In the left panel, we show the analytical estimate for the electric field anistopies contribution to the GW spectrum and in the right panel show the analytical estimate for the magnetic fields anisotropic stresses contribution after reheating. The first and second vertical gridlines correspond to the frequencies $\nu_H$(frequency corresponding to wavenumber $k_H$) and $\nu_0$(frequency corresponding to wavenumber $k_0$) respectively. We plotted the analytical estimates for the modes $\nu\leq\nu_0$. Although the approximation used in deriving analytical estimate fails for $\nu > \nu_H$, we extrapolate the analytical estimates until the the modes $\nu<\nu_0$.}
\label{gwnh}
\end{figure*}
For $k< k_H$ the expression for $d\Omega_{GW}/d \ln k \big|_0$ is given by,
\begin{align}\label{analytical}
\frac{d\Omega_{GW}}{d \ln k}\Bigg|_0&=\frac{7 \Omega_R}{5} \left(\frac{k}{k_0}\right)^3 \left(\frac{D_2}{\tilde{\rho}+\tilde{p}}\right)^2\left(\frac{64 \beta^2}{(1-4 \beta )^2 (8 \beta +1)^2}\right)+c \Omega_R \left(\frac{D_1}{\tilde{\rho}+\tilde{p}}\right)^2 \left(\frac{k}{k_0}\right)^3.
\end{align}
where,
\begin{align}
    k_0=1.72\times 10^{9} \beta \left(\frac{g_R}{106.75}\right)^{1/6} \frac{T_R}{100 GeV}Mpc^{-1}
\end{align}
and the corresponding frequency 
\begin{align}
    \nu_0=\frac{k_0}{2 \pi}=2.7\times 10^{-6} \beta \left(\frac{g_R}{106.75}\right)^{1/6} \frac{T_R}{100 GeV} Hz
\end{align}
[See Appendix \ref{3} for details]. Here $g_R$ is the relativistic degree of freedom at the epoch of reheating. In the above expression, the first part represents the analytical estimate of the contribution from EM field anisotropic stresses before reheating and the second part represents that from the magnetic field anisotropic stresses after reheating. This is compared with the
results from numerical integration for the low wavenumbers.  Although the approximation used in deriving analytical estimate fails for $\nu > \nu_H$, we extrapolate the analytical estimates until the the modes $\nu<\nu_0$ and show the comparison with the numerical result in \fig{gwnh}. An estimate of the GW background 
amplitude can be obtained using \eqref{analytical}
for the mode $k=k_0$. Adopting
$T_R=100$ GeV, $\Omega_R=9.24 \times 10^{-5}$, $c=0.31$ and $\beta=6.37$ 
(if the ratio of EM energy density to the background energy density 
($\epsilon$) is one at reheating) gives,
\begin{align}
\frac{d\Omega_{GW}}{d \ln k}\Bigg|_{0(k=k_H)}&\approx 1.7\times 10^{-8} ~~~\text{and}~~~~\frac{d\Omega_{GW}}{d \ln k}\Bigg|_{0(k=k_0)}\approx 4.3\times 10^{-6}.
\end{align}
This amplitude decreases roughly as $D_1^2 \propto \epsilon^2$, and
so is approximately $10^{-4}$ times smaller for $\epsilon=10^{-2}$. This is an approximate estimate as $\beta$ also changes slowly with $\epsilon$. Note that our analytical estimate for the GW energy spectrum differs by a factor of 2 for $k=k_H$ and 5 for $k=k_0$ by the numerical estimate. Our primary aim to give the analytical estimate is to know the spectral nature which matches well with the numerical estimate within the region where the approximation made in analytical estimate is valid as shown in \fig{gwnh}.

\begin{figure*}
\includegraphics[scale=0.4]{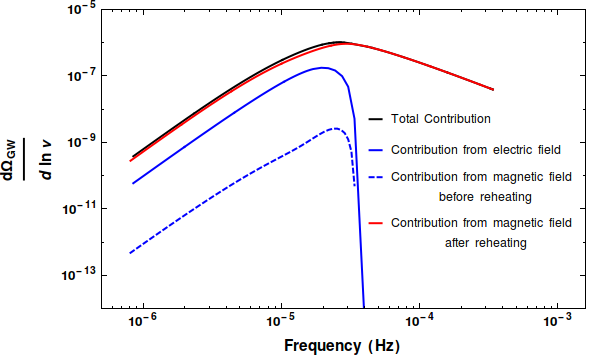}
\includegraphics[scale=0.4]{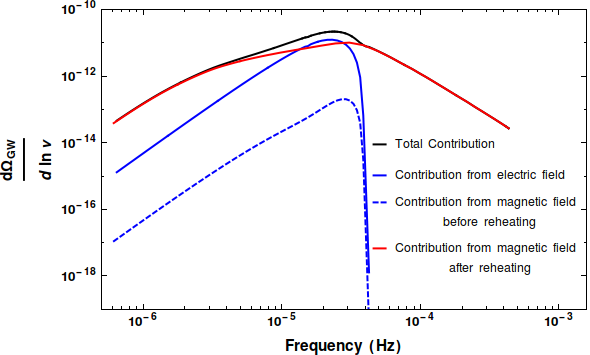}
  \caption{In this figure we plot the different contribution to the GW energy spectrum generated from the EM field anisotropic stresses (non helical case). In the left and right panel, we assume $\epsilon=1$ and $\epsilon=10^{-2}$, respectively. The blue and the dashed blue lines, respectively, represent the contribution to the GW energy spectrum from electric and magnetic fields anisotropic stresses before reheating. The red curve represent the contribution from the magnetic field anisotropic stresses after reheating and the black curve represent the sum of all these contributions.}
\label{gwnh2}
\end{figure*}  

\subsubsection{Numerical results for GW spectrum}

We calculate the GW spectrum for different reheating temperatures $T_R$ 
and different fractions ($\epsilon$) of the EM field energy density 
to the background energy density at the time of reheating.
Our model of magnetic field generation during inflation, requires reheating
to be below an energy scale of $5000$ GeV to satisfy the constraints from 
the $\gamma$-ray observations \cite{sharma2017} which changes to the value $500$ GeV 
in case of helical nature of EM field \cite{sharmahelical}. At the same time, it should be above $5$ MeV to
account for Big Bang Nucleosynthesis \cite{bbn}. We therefore give results for some representative values of $T_R$
which lie in this range. Each wavenumber $k$ is also be converted into 
the frequency $\nu$ of the GW using $\nu= k c/2 \pi$.

In Fig.~(\ref{gwnh2}), we have shown the different contribution to the GW energy spectrum for $\epsilon=1$ and $\epsilon=10^{-2}$ assuming $T_R=100$ GeV. The blue and dashed blue curve shown the contribution from the electric and magnetic fields spectrum before reheating, respectively. The red curve shows the contribution from the magnetic field spectrum after reheating. As is evident from this figure, the main contribution to the GW energy spectrum comes from the magnetic field anisotropic stresses after reheating for $\epsilon=1$. However, for $\epsilon=10^{-2}$, the contribution from electric field anisotropic stresses dominate around the peak of the total GW spectrum elsewhere it is dominated by the contribution from magnetic field anisotropic stresses after reheating. This leads to an extra bump type feature around the peak in the resultant GW energy spectrum for $\epsilon=10^{-2}$.

\begin{figure*}
\includegraphics[scale=0.47]{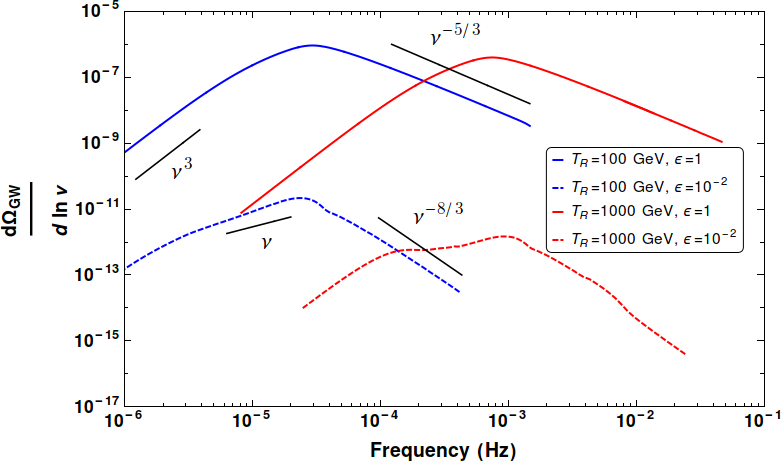}
\includegraphics[scale=0.5]{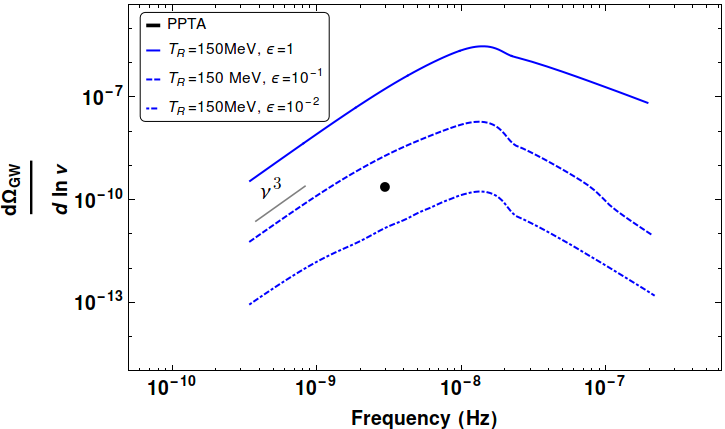}
  \caption{In this figure we plot the GW energy spectrum generated from the EM field anisotropic stresses (non helical case). In the upper panel, we plot GW energy spectrum for the reheating scale $T_R=100$ GeV and $T_R=1000$ GeV and also for the different fraction ($\epsilon$) of EM field energy density to the background energy density at reheating.
  In the lower panel, we plot the GW spectrum for the reheating scale at $T_R=150$ MeV. The black dot point in the lower panel of the figure represent the limit on the GW energy spectrum at the nanohertz scale obtained from the Parkes pulsar timing array (PPTA) \cite{ppta}.}
\label{gwnh1}
\end{figure*}

In the upper panel of Fig.~(\ref{gwnh1}), we plot GW energy spectrum 
for the reheating scale $T_R=100$ GeV and $T_R=1000$ GeV and 
also for $\epsilon=1$ and $\epsilon=10^{-2}$.
The peak of the GW spectrum lies approximately around the peak of the 
EM field spectrum at reheating. 
Note that the electric and magnetic field spectra peak at the frequency corresponding to a wavenumber which is $\beta$ times of the horizon wavenumber at reheating and this $\propto 1/T_R$. This relation, however, is approximate since $\beta$ also depends mildly on the value of $T_R$. Therefore 
the frequency at which GW spectrum has its peak, has a roughly linear behaviour with the reheating temperature, 
$T_R$. It is also weakly dependent on $\epsilon$ corresponding to the same $T_R$ due to change in the value of $\beta$ for different $\epsilon$. 
The peak value of the GW spectrum is $d\Omega_{GW}/d\ln(k) \approx 9.6 \times 10^{-7}$ 
at the frequency $30~ \mu$Hz for $T_R=100$ GeV 
and $4.1 \times 10^{-7}$ at the frequency $1$ mHz for $T_R=1000$ GeV assuming $\epsilon=1$ 
for both the cases. For $\epsilon=10^{-2}$, the peak value of the GW spectrum changes to 
$2.0\times 10^{-11}$ for $T_R=100$ GeV and to $1.5 \times 10^{-12} $ for $T_R=1000$ GeV, respectively. 
The approximate dependence $\Omega_{GW} \propto \epsilon^2$ for a given $T_R$ is because
the amplitude of tensor metric perturbations depend on the amplitude of the anisotropic stress 
of the EM field (which is $\propto \epsilon$)
and $\Omega_{GW}$ depends quadratically on these metric perturbations.
For the modes $k \leq k_H$ (the mode where GW energy spectrum peaks), our analytical estimate in Eq.~\eqref{analytical} suggests that
the spectrum is proportional to $k^3$. 
As is evident from Fig.~\ref{gwnh2}, for the modes $k>k_{peak}$, the GW energy spectrum is proportional to $k^{-5/3}$ for $\epsilon=1$ and $k^{-8/3}$ for $\epsilon=10^{-2}$. The slope $k^{-8/3}$ for the case $\epsilon=10^{-2}$ matches with the result obtained from the numerical simulation in \citet{axelgw}.  The frequency at which the GW energy spectrum peaks, $\nu_{peak}\approx 2~\nu_0$ ($\nu_0$ is the frequency corresponding to the wavenumber $k_0=k_p(\eta_R)$) for any $T_R$ and $\epsilon$.

We show the predicted GW spectrum for the lower reheating scale at $T_R=150$ MeV
in the lower panel of Fig.~(\ref{gwnh1}). For the modes $k \leq k_{peak}$, the spectrum is proportional 
to $k^3$ similar to other reheating scales. For this case, the GW spectrum has the peak value 
$3.1 \times 10^{-6}$ at the frequency $1.6 \times 10^{-8}$ Hz. 
Present limits on the GW spectrum at nanohertz frequencies
are obtained from Parkes pulasar timing array (PPTA) \cite{ppta}. This is shown as a
black dot in the lower panel of the figure.
Since PPTA does not detect any 
GW with this sensitivity, from Fig.(\ref{gwnh1}) we conclude that for 
$T_R=150$ MeV, $\epsilon < 10^{-1}$. This limit will become even stronger for those 
scenarios in which reheating is below 150 MeV and as the pulsar timing
array limits improve in the future.

\subsection{Gravitational waves energy spectrum for helical EM fields}
\label{helical}

If we generalise the Ratra model of inflationary magnetogenesis and add a parity breaking term $(f^2 F_{\mu \nu}\tilde{F}^{\mu \nu})$ to the Lagrangian density, the generated magnetic field is almost fully helical. Here $\tilde{F}^{\mu \nu}$ is the dual tensor of the EM tensor. In the standard electromagnetism with $f$ constant, this term is a total divergence term and does not contribute to the evolution of the electromagnetic field. However when $f$ is time dependent and conformal invariance of the EM theory is broken, this term contributes to the evolution of the EM field. It introduces a mixing between the two vector potential modes in the linear polarisation basis. In terms of the helicity basis (or circular polarisation basis), left and right circular polarization
modes decouple and
satisfy different evolution equations. It turns out that the amplitude of one of the heilcal mode is larger than the other at the end of generation era and the net generated magnetic field is of helical nature. In our model \cite{sharmahelical}, the generated magnetic field is almost fully helical. This results in the generated stochastic GW to be predominantly
circularly polarized. The calculations of the GW background which is generated by such a field is on similar lines as for the non-helical case. Hence, we only give the results of the GW, magnetic and electric energy spectrum.

\begin{figure*}
\includegraphics[scale=0.5]{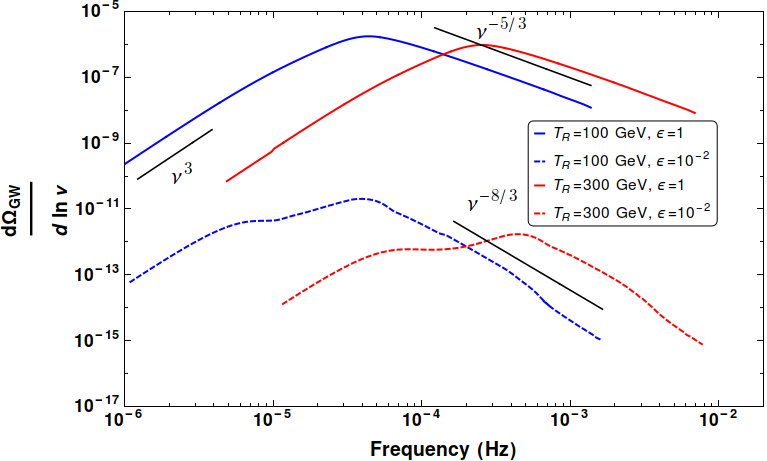}
  \caption{In this figure we plot the GW energy spectrum generated from the EM field anisotropic stresses (helical case).}
\label{gwh}
\end{figure*}

For this case $|\Pi^\aa|^2(k,\eta,\eta')$ is given by,
\begin{equation} \label{piinhelical1}
|\Pi^-|^2 (k,\eta,\eta') = \left\{
        \begin{array}{ll}
           \frac{a_R}{a(\eta)}\frac{a_R}{a(\eta')}\left(\frac{1}{\tilde{\rho}+\tilde{p}}\right)^2\frac{1}{2}\Big(g_B(k,\eta,\eta')+g_E(k,\eta,\eta')+h_B(k,\eta,\eta')+h_E(k,\eta,\eta')\Big),&  \eta, \eta'\leq\eta_R\\
          \left(\frac{1}{\tilde{\rho}+\tilde{p}}\right)^2 \frac{1}{2}\Big(g_B(k,\eta,\eta')+h_B(k,\eta,\eta'),&  \eta, \eta'\geq\eta_R
        \end{array}
    \right.
\end{equation}
\begin{equation} \label{piinhelical2}
|\Pi^+|^2 (k,\eta,\eta') = \left\{
        \begin{array}{ll}
           \frac{a_R}{a(\eta)}\frac{a_R}{a(\eta')}\left(\frac{1}{\tilde{\rho}+\tilde{p}}\right)^2\frac{1}{2}\Big(g_B(k,\eta,\eta')+g_E(k,\eta,\eta')-h_B(k,\eta,\eta')-h_E(k,\eta,\eta')\Big),&  \eta, \eta'\leq\eta_R\\
          \left(\frac{1}{\tilde{\rho}+\tilde{p}}\right)^2 \frac{1}{2}\Big(g_B(k,\eta,\eta')-h_B(k,\eta,\eta'),&  \eta, \eta'\geq\eta_R
        \end{array}
    \right.
\end{equation}
The above expressions imply that $|\Pi^-|^2$ is larger than $|\Pi^+|^2$ by an amount $h_B(k,\eta,\eta')+h_E(k,\eta,\eta')$.
Adding Eq.(\ref{piinhelical1}) and Eq. (\ref{piinhelical2}), we get
\begin{equation} \label{piinhelical}
|\Pi^-|^2 (k,\eta,\eta')+|\Pi^+|^2 (k,\eta,\eta') = \left\{
        \begin{array}{ll}
           \frac{a_R}{a(\eta)}\frac{a_R}{a(\eta')}\left(\frac{1}{\tilde{\rho}+\tilde{p}}\right)^2\frac{1}{2}\Big(g_B(k,\eta,\eta')+g_E(k,\eta,\eta')\Big),&  \eta, \eta'\leq\eta_R\\
          \left(\frac{1}{\tilde{\rho}+\tilde{p}}\right)^2 \frac{1}{2}\Big(g_B(k,\eta,\eta'),&  \eta, \eta'\geq\eta_R
        \end{array}
    \right.
\end{equation}
In the above expression tilde over quantities represents their comoving values. To estimate $g_B$ and $g_E$, we need to know the electric and magnetic energy density in the matter dominated era after inflation and their evolution after reheating. The comoving spectral electric and magnetic field energy density in the matter dominated era before reheating ($\eta\leq \eta_R$) for this case are [\citet{sharmahelical}],
\begin{equation}\label{rhobinhelical}
\frac{d \tilde{\rho}_B(k, \eta)}{d \ln k} = \left\{
        \begin{array}{ll}
           D_{1h} \left(\frac{k}{\kp(\eta)}\right)^4\left(\frac{\eta+\eta_R}{2\eta_R}\right)^{8 \beta+2},  k\leq \kp(\eta)\\
           D_{1h} \left(\frac{k}{\kp(\eta)}\right)^4\left(\frac{\eta_k+\eta_R}{2\eta_R}\right)^{8 \beta+2},k\geq \kp(\eta)
        \end{array}
    \right.
\end{equation}
\begin{equation}\label{rhoeinhelical}
    \frac{d \tilde{\rho}_E(k,\eta)}{d \ln k}= \left\{
    \begin{array}{ll}
D_{2h} \left(\frac{k}{\kp(\eta)}\right)^2 \left(\frac{\eta+\eta_R}{2\eta_R}\right)^{8 \beta},k\leq \kp(\eta)\\
 D_{2h} \left(\frac{k}{\kp(\eta)}\right)^4\left(\frac{\eta_k+\eta_R}{2\eta_R}\right)^{8 \beta},k\geq \kp(\eta)
\end{array}
\right.
\end{equation}
Here $D_{1h}$ and $D_{2h}$ are the amplitudes of spectral magnetic and electric energy densities at $k_0=k_p(\eta_R)$ respectively. The values of $D_{1h}$ and $D_{2h}$ depend on the fraction of electromagnetic energy density to background energy density at reheating.
We again consider GW production in two scenario on the basis of evolution of the magnetic field energy density after reheating.

After reheating, non linear processing (as in non helical case) of magnetic field spectrum takes place. However, the magnetic field energy density decays at a rate slower as compared to the case of non helical magnetic field because of the helicity conservation \cite{kandu2016,jedamzik, axel2001}. The evolution of the magnetic field for $\eta \geq \eta_{R}$ is given by,
\begin{equation}\label{rhononlinearinhelical}
    \frac{d \tilde{\rho}_B(k)}{d \ln k}= \left\{
        \begin{array}{ll}
    D_{1h} \left(\frac{k}{k_{NL}(\eta)}\right)^4 \left(\frac{\eta}{\eta_R}\right)^{-2/3}, k\leq k_{NL}(\eta)\\
    D_{1h} \left(\frac{k}{k_{NL}(\eta)}\right)^{-2/3} \left(\frac{\eta}{\eta_R}\right)^{-2/3},  k_{\nu}\geq k\geq k_{NL}(\eta)
\end{array}
    \right.
\end{equation}
Here,
\begin{align}
k_{NL}(\eta) &= k_0 \left(\frac{\eta}{\eta_R}\right)^{-2/3}
\end{align}

By substituting Eq.\eqref{rhobinhelical},\eqref{rhoeinhelical} and \eqref{rhononlinearinhelical} in Eq.\eqref{piinhelical}, we estimate $\sum_\aa |\Pi^\aa|^2$ which we further substitute in Eq. \eqref{hijexp} to calculate $\sum_\aa |dh^{\aa}/d x|^2$. After substituting $\sum_\aa |dh^{\aa}/d x|^2$ in Eq. \eqref{omegainhp} we calculate GW energy density spectrum for this case and the results are shown in Fig.(\ref{gwh}). The peak value of the GW spectrum is $1.8 \times 10^{-6}$ at the frequency $40~\mu$Hz for $T_R=100$ GeV and $9.5 \times 10^{-7}$ at the frequency $0.25$ mHz for $T_R=300$ GeV assuming $\epsilon=1$ for both the cases. The peak values of the GW spectrum in this case is approximately twice the value in the case of non-helical magnetic field. This is due to the fact that there is extra contribution to the GW spectrum which comes from the antisymmetric term in the two point correlation of the electric and magnetic field spectrum compare to the non-helical case.

For the helical nature of the EM field, the generated GW spectrum is circularly polarised. Eq.\eqref{piinhelical1} and Eq.\eqref{piinhelical2} suggest that the negatively polarised (-) mode dominates over the positively polarised (+) mode by an amount $h_E(k, \eta, \eta')+h_B(k, \eta, \eta')$.
However the spectrum is unpolarised for the case when EM field is of non-helical nature since the contribution of both the modes is equal as can be seen from Eq.\eqref{piinnonhelical}. The sign of the GW polarisation that dominates (+ or -), depends on the relative sign of the parity breaking term ($F_{\mu \nu} \tilde{F}^{\mu \nu}$) to the standard term ($F_{\mu \nu} F^{\mu \nu}$) in the EM field Lagrangian. In our analysis, we have taken both the terms with the same sign.
 
\section{Detection of the generated GW spectrum with the LISA}\label{detectionwlisa}
We now discuss the prospects of detection of the GW spectrum generated in our model with the LISA. For this, we calculate the signal to noise ratio (SNR) using the following definition \cite{caprini2019},
\begin{align}
    \text{SNR}\equiv\sqrt{T\int_{\nu_{min}}^{\nu_{max}} d\nu \left(\frac{d \Omega_{GW}}{d \ln \nu}\middle/\frac{d \Omega_n}{d \ln \nu}\right)^2}.
\end{align}
In the above expression, $d \Omega_n/d \ln \nu=(4 \pi^2/3 H_0^2)\nu^3 S_n(\nu)$ where $S_n(\nu)$ is the strain sensitivity of the LISA detector and $T$ is the mission duration. The integration limits, $\nu_{min}$ and $\nu_{max}$, denote the minimal and maximal frequencies accessible at
the LISA detector respectively.
It is convenient to express the SNR as,
\begin{align}
    \text{SNR}=\sqrt{\int_{\nu_{min}}^{\nu_{max}} d \ln \nu \left(\sqrt{\nu T}\frac{d \Omega_{GW}}{d \ln \nu}\middle/\frac{d \Omega_n}{d \ln \nu}\right)^2}\equiv\sqrt{\int_{\nu_{min}}^{\nu_{max}} d \ln \nu        ~(\text{SNR}(\nu))^2},
\end{align}
where
\begin{align}\label{snrf}
  \text{SNR}(\nu)= \sqrt{\nu T}\left(\frac{d \Omega_{GW}}{d \ln \nu}\middle/\frac{d \Omega_n}{d \ln \nu}\right).
\end{align}

\begin{figure*}
\includegraphics[scale=0.355]{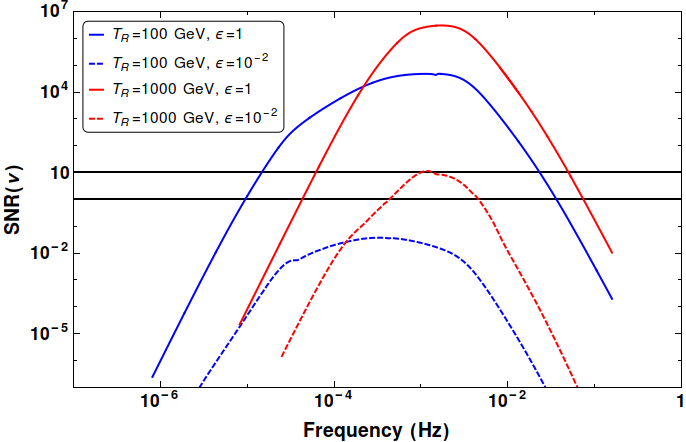}
\includegraphics[scale=0.345]{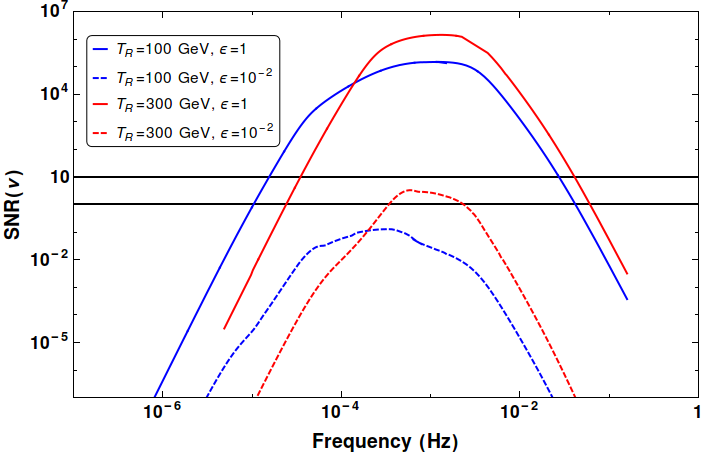}
  \caption{In this figure, we plot SNR($\nu$) defined in Eq.~\eqref{snrf} vs frequency for non helical and helical nature of the EM field. In the left panel, we plot SNR($\nu$) for the reheating scale $T_R=100$ GeV and $T_R=1000$ GeV and also for the different fraction ($\epsilon$) of EM field energy density to the background energy density for non helical nature of EM field. In the right panel, we plot SNR($\nu$) for the reheating scale $T_R=100$ GeV and $T_R=300$ GeV and also for the different fraction ($\epsilon$) of EM field energy density to the background energy density for helical nature of EM field. The lower and upper black horizontal lines represents SNR($\nu$)=1 and SNR($\nu$)=10, respectively. For these plots, we take $T=3$ years.}
\label{lisadetect}
\end{figure*}
Thus the square of the SNR($\nu$) provides the the contribution to the value of square of the SNR per logarithmic frequency interval. To calculate the SNR for different reheating temperature $T_R$, we used the strain sensitivity $S_n(f)$ of a single channel of the LISA detector given in Eq. (2.4) in \citet{caprini2019}.

In \fig{lisadetect}, we plot SNR($\nu$) with frequnecy ($\nu$). This figure shows the sensitivity of the LISA detector for our generated GW spectrum in different frequency bands.  In the Fig.~(\ref{lisadetect}), the lower and upper black horizontal lines represents SNR($\nu$)=1 and SNR($\nu$)=10, respectively.
As evident from Fig.(\ref{lisadetect}), the generated GW spectrum lies 
in the sensitivity range of LISA for 
our magnetic field 
generation models, in which $\epsilon$ is above a threshold value which depends on $T_R$. 
For $\epsilon=10^{-2}$, the GW spectrum generated in our model can be detected with an SNR=10 for $T_R=1000$ GeV in non helical case and SNR=3 for $T_R=300$ GeV for helical case. In these estimates, we take $T=3$ years. The value of SNR is even higher for large value of $\epsilon$. Although, we show explicitly
the GW spectrum only for $T_R=100$ GeV and $T_R=1000$ GeV for non helical case and $T_R=100$ GeV and $T_R=300$ GeV for helical case, 
the nature of the spectrum is qualitatively similar for other values of $T_R$.

\section{Discussion and Conclusion}\label{conclusion}

Origin of large scale magnetic fields in the universe
is a subject of intense study. An intriguing possibility is its generation
during inflation
with the model suggested by \citet{turner-widrow} and \citet{ratra} being a popular scenario. 
However this model potentially suffers from what are known as the
strong coupling and back-reaction problems. In our earlier studies (\citet{sharma2017,sharmahelical}), 
we suggested a model to address these issues. 
We showed
that for a certain range of inflationary and reheating scales, 
it is possible to generate magnetic fields with the strengths of astrophysical interest while at the same time addressing all the difficulties raised regarding the models of Ref. \cite{turner-widrow, ratra}. Our models required a low energy scale of inflation
and reheating, with $T_R < 5000$ GeV for magnetogenesis
scenarios which generate nonhelical fields and $T_R < 500$ GeV for helical
models \cite{sharma2017,sharmahelical}. They also predicted a blue spectrum peaked around the
horizon scale of reheating with EM fields a
significant fraction of the cosmological energy density at that epoch.
These EM fields have non zero anisotropic stresses 
which can source the production of a stochastic background of GW. 

Here, we have therefore calculated the spectrum of the resulting stochastic GW background. Our aim was also
to probe and constrain such models of
inflationary magnetogensis by examining whether the predicted
GW spectrum can be detected
by upcoming space mission LISA or PTA experiments.
We obtained the GW spectrum for both 
magnetogenesis models where the generated EM fields are
non-helical 
or models which resulted in helical magnetic fields.
An analytical estimate of the GW spectrum for low wavenumbers and non-helical fields is given by \eq{analytical}. The results of 
numerical integration are given in \fig{gwnh1} when the source of GW are
nonhelical primordial EM fields and in \fig{gwh} for helical primordial fields, 
for some representative values of $T_R=100$ GeV, $T_R=300$ GeV, $T_R=1000$ GeV and $T_R=150$ MeV. 
Similar results can be obtained for other values
of $T_R$. 
We also estimated these GW spectra for different fraction ($\epsilon$) of EM field 
energy to the background energy density at reheating for each temperature scale. 

The generated GW spectrum rises with wavenumber $k$ as $d \Omega_{GW}/d \ln(k) \propto k^3$,
at low wavenumbers. It remains almost $k^3$ until the wavenumber $k=k_{peak}$ for the fraction of EM energy density to the background energy density at reheating, $\epsilon=1$. However, for $\epsilon=10^{-2}$, it changes to a spectrum $\propto k$. The GW spectrum then falls for the modes $k>k_{peak}$ as $d \Omega_{GW}/d \ln(k) \propto k^{-5/3}$ for $\epsilon=1$ and $d \Omega_{GW}/d \ln(k) \propto k^{-8/3}$ for $\epsilon=10^{-2}$. This change in the slope of the spectrum for different $\epsilon$ could arise due to the fact that the turbulence correlation time is longer for a smaller $\epsilon$.
For $\epsilon=1$, the peak value of the generated GW spectrum, $d\Omega_{GW}/d \ln k \sim 4.1 \times 10^{-7} $ 
for the non-helical case at $T_R=1000$ GeV and $d\Omega_{GW}/d \ln k \sim 9.5 \times 10^{-7} $ for the helical case at $T_R=300$ GeV. The amplitude at the peak value decreases approximately as
$\epsilon^2$. Note that each wavenumber $k$ can be converted into the frequency $\nu$ of the GW using $\nu = k c/2\pi$. The corresponding frequencies for these $T_R$ is in mHz range where LISA is sensitive to detect a GW signal. The amplitude is similar for a lower value of $T_R$, but the frequency
at which the GW spectrum peaks decreases as $\nu_{peak} \propto T_R$ approximately.

Our results show that 
the strength of the generated GW in both nonhelical and helical cases are of similar order. 
However, in the case of helical EM fields, the generated GW spectrum is circularly polarised while it is unpolarised
when the generated EM fields are
nonhelical.
All the scenarios in which reheating is above the 100 GeV scale produce 
a GW energy spectrum which lie in the sensitivity range of LISA 
provided that the fraction of EM field energy density to the background 
energy density ($\epsilon$) is above a threshold value of order $10^{-2}$. 
For $\epsilon=10^{-2}$, the GW spectrum can be detected with an SNR=10 for $T_R=1000$ GeV in non helical case and SNR=3 for $T_R=300$ GeV for helical case. For these estimates of SNR, we assumed $T=3$ years. The large value of $\epsilon$ gives larger value of SNR. For lower reheating temperature $T_R = 150$ MeV, the peak frequency
shifts to $20$ nano Hertz, where PTA experiments are more relevant. The current
limits from PPTA constrain $\epsilon < 10^{-1}$ in this case.

Stochastic GW at these frequencies can also result from first order phase transitions
at the corresponding temperatures. The inflationary models considered here can however be
distinguished from the signals arising in such first order phase transitions
due to the following.  
As evident from the right panel of \fig{gwnh2}, there is a bump like feature around $\nu_{peak}$ in the resultant GW energy spectrum for the more realistic case of $\epsilon=10^{-2}$. 
This happens due to the fact that the contribution from the electric field anisotropic stresses before reheating dominates over the contribution from the magnetic field anisotropic stresses around the peak value of the spectrum and the total spectrum gets an additional contribution aroung the peak value.
The GW spectrum generated during phase transition is also proportional 
to $k^3$ for the modes below the peak value and has another branch for the 
modes above the peak value developed due to Kolomogorov branch of the decaying magnetohydrodynamic turbulence.
However, in the phase transition generated spectrum, there is no bump in these 
two branches around the peak value like in our case \citep{axelgw}. 
This feature is unique to our model of inflationary magnetogenesis. Another distinguishing feature of our model is the possibility of obtaining an almost fully circularly polarised stochastic GW background.
A possible detection of GW by LISA or by PTA, with the features predicted here will provide an important probe of several models
of inflationary magnetogenesis.

\section*{Acknowledgments}
RS and TRS acknowledge the facilities at I.C.A.R.D., University of Delhi. RS also acknowledges C.S.I.R., India for the financial support through grant 09/045(1343)/2014-EMR-I. TRS acknowledges the project grant from SERB EMR/2016/002286. Authors would like to thank Axel Brandenburg, Sukanta Bose and Sanjit Mitra for useful discussions and helping out with LISA sensitivity curve. RS thanks Axel Brandenburg for hosting him at Nordita for the event "Gravitational Waves from the early universe", during which time a part of this work was done.
\appendix
\section{Four point correlation function}
To estimate the expression in Eq.~\eqref{coreqn}, we need to calculate the four point correlation function of electric and magnetic field (for both nonhelical and helical case). Using the gaussian nature of the EM field, we can express the four point correlation function, $\langle\tilde{B}_{m}(\vec{q},\eta) \tilde{B}^*_{n}(\vec{q}-\vec{k},\eta)  \tilde{B}^{*a}(\vec{r},\eta') \tilde{B}^{b}(\vec{r}-\vec{k'},\eta')  \rangle$, in terms of the two point correlation function as,
\begin{align}
\langle\tilde{B}_{m}(\vec{q},\eta) \tilde{B}^*_{n}(\vec{q}-\vec{k},\eta)  &\tilde{B}^{*a}(\vec{r},\eta') \tilde{B}^{b}(\vec{r}-\vec{k'},\eta')  \rangle\nonumber\\
=&\langle\tilde{B}_{m}(\vec{q},\eta) \tilde{B}^*_{n}(\vec{q}-\vec{k},\eta)\rangle \langle  \tilde{B}^{*a}(\vec{r},\eta') \tilde{B}^{b}(\vec{r}-\vec{k'},\eta')  \rangle\nonumber\\
&+\langle\tilde{B}_{m}(\vec{q},\eta) \tilde{B}^{*a}(\vec{r},\eta')\rangle \langle \tilde{B}^*_{n}(\vec{q}-\vec{k},\eta)  \tilde{B}^{b}(\vec{r}-\vec{k'},\eta')  \rangle\nonumber\\
&+  \langle\tilde{B}_{m}(\vec{q},\eta)\tilde{B}^{b}(\vec{r}-\vec{k'},\eta')\rangle \langle \tilde{B}^*_{n}(\vec{q}-\vec{k},\eta)  \tilde{B}^{*a}(\vec{r},\eta')   \rangle
\end{align}
We can further divide the analysis into two parts on the basis of helical and nonhelical nature of EM field. 
\subsection{Non-helical EM field}\label{appendixa}
We use Eq.~\eqref{tpce} to represent two point correlation function in terms of the power spectra ($P_{SB}$) and unequal time correlation ($C_B$) of the magnetic field.
\begin{align}
&\langle\tilde{B}_{m}(\vec{q},\eta) \tilde{B}^*_{n}(\vec{q}-\vec{k},\eta)  \tilde{B}^{*a}(\vec{r},\eta') \tilde{B}^{b}(\vec{r}-\vec{k'},\eta')  \rangle\nonumber\\
=&(2\pi)^6\Bigg((\delta_{mn}-\hat{q}_m \hat{q}_n) \delta(\vec{k})(\delta^{ab}-\hat{r}^a \hat{r}^b) \delta(\vec{k'}) P_{SB}(q,\eta) P_{SB}(r,\eta')+\Big((\delta_{m}^a-\hat{q}^a \hat{q}_m) \delta(\vec{q}-\vec{r})\nonumber\\
&(\delta^b_n-(\widehat{k-q})^b (\widehat{k-q})_n)\delta(\vec{q}-\vec{r}-\vec{k}+\vec{k'}) P_{SB}(q,\eta) P_{SB} (|\vec{k}-\vec{q}|,\eta)+(\delta_{m}^{b}-\hat{q}^b \hat{q}_m) \delta(\vec{q}+\vec{r}-\vec{k'})\nonumber\\
&(\delta^{a}_{n}-(\widehat{k-q})^a (\widehat{k-q})_n) \delta(\vec{r}+\vec{q}-\vec{k})P_{SB}(q,\eta) P_{SB}(|\vec{k}-\vec{q}|,\eta)\Big)C_B(q,\eta,\eta') C_B(|\vec{k}-\vec{q}|,\eta,\eta')\Bigg)
\end{align}
In the above expression, we use Eq.~\eqref{dtcf} to represent unequal time correlation function in terms of equal time correlation function. After taking the projection to calculate the transverse traceless part and integrating the above equation with respect to $r$, we get
\begin{align}
&\int \frac{d^3 r}{(2 \pi)^3}P^{mn}_{ij}(\vec{k}) P^{ij}_{ab}(\vec{k'})\langle\tilde{B}_{m}(\vec{q},\eta) \tilde{B}^*_{n}(\vec{q}-\vec{k},\eta)  \tilde{B}^{*a}(\vec{r},\eta') \tilde{B}^{b}(\vec{r}-\vec{k'},\eta')  \rangle\nonumber\\
=&(2\pi)^3P^{mn}_{ij}(\vec{k}) P^{ij}_{ab}(\vec{k'})\Bigg((\delta_{m}^a-\hat{q}^a \hat{q}_m)(\delta^b_n-(\widehat{k-q})^b (\widehat{k-q})_n)+(\delta_{m}^{b}-\hat{q}^b \hat{q}_m)(\delta^{a}_{n}-\nonumber\\
&(\widehat{k-q})^a (\widehat{k-q})_n)\Bigg) P_{SB}(q,\eta) P_{SB}(|\vec{k}-\vec{q}|,\eta)C_B(q,\eta,\eta') C_B(|\vec{k}-\vec{q}|,\eta,\eta')\delta(\vec{k}-\vec{k'}) \nonumber\\
=&(2\pi)^3 (1+\gamma^2+\beta^2+\gamma^2\beta^2)P_{SB}(q,\eta) P_{SB}(|\vec{k}-\vec{q}|,\eta))C_B(q,\eta,\eta') C_B(|\vec{k}-\vec{q}|,\eta,\eta')\delta(\vec{k}-\vec{k'}) 
\end{align}
Here, $\gamma=\hat{k}\cdot \hat{q}$ and $\beta=\hat{k}\cdot \widehat{k-q}$.
\subsection{Helical EM field}\label{appendixb}
In this case, we use Eq.~\eqref{tpceh} to represent two point correlation function in terms of the symmetric ($P_{SB}$) and anti-symmetric ($P_{AB}$) part of the power spectrum of the magnetic field. Since Eq.~\eqref{tpceh} has a term containing $P_{AB}$ unlike the nonhelical case, we also have contribution from this term in the four point correlation depicted as follows,
\begin{align}
&\int \frac{d^3 r}{(2 \pi)^3}P^{mn}_{ij}(\vec{k}) P^{ij}_{ab}(\vec{k'})\langle\tilde{B}_{m}(\vec{q},\eta) \tilde{B}^*_{n}(\vec{q}-\vec{k},\eta)  \tilde{B}^{*a}(\vec{r},\eta') \tilde{B}^{b}(\vec{r}-\vec{k'},\eta')  \rangle\nonumber\\
=&(2\pi)^3 \left((1+\gamma^2+\beta^2+\gamma^2\beta^2)P_{SB}(q,\eta) P_{SB}(|\vec{k}-\vec{q}|,\eta))+4\gamma \beta P_{AB}(q,\eta) P_{AB}(|\vec{k}-\vec{q}|,\eta))\right)\nonumber\\
&C_B(q,\eta,\eta') C_B(|\vec{k}-\vec{q}|,\eta,\eta')\delta(\vec{k}-\vec{k'}) 
\end{align}
As we neglect the charge density during the matter dominated era before reheating, similar expression for the four point correlation function for electric field can be obtained by replacing the magnetic field power spectrum ($P_{SB}$) with the electric field power spectrum ($P_{SE}$). By substituting the above expressions in Eq.~\eqref{coreqn}, we obtain Eq.~\eqref{piijnh} and Eq.~\eqref{piijh}, respectively.
\section{Calculation to estimate $d_1$ and $d_2$}\label{hcalculation}
To calculate $d_1$ and $d_2$ given in Eq.~\eqref{maiequation}, we match $h^{\aa}$ and its derivative at the epoch of reheating ($\eta=\eta_R$). The matching relations are,
\begin{align}
h^\aa(\yy_R)&=h^\aa(x_R)\nonumber\\
\frac{d h^\aa(\yy)}{d\yy}\Big|_{\yy=\yy_R}&=\frac{d h^\aa(x)}{d x}\Big|_{x=x_R}
\end{align}
Here, $\yy_R=2 k \eta_R$ and $x_R=k \eta_R$ which implies $\yy_R=2 x_R$.
Using Eq.~\eqref{hijinmd} and Eq.~\eqref{maiequation}, the above two conditions imply,
\begin{eqnarray}
d_1&=&\frac{1}{32 x_R^3}\Bigg( \left(\int_{\yy_i}^{2 x_R} \frac{12 \Pi^\aa (\yy_1) (\sin \yy_1-\yy_1 \cos \yy_1)}{\yy_1} d\yy_1\right)\Big(-8 x_R^2 \cos x_R+4 x_R \sin x_R\nonumber\\
&&+\cos x_R+\cos 3 x_R\Big)+(8 x_R^2 \sin x_R+4 x_R \cos x_R-\sin x_R-\sin 3 x_R)\nonumber\\
&& \left(\int_{\yy_i}^{2 x_R} \frac{12 \Pi^\aa (\yy_1) (\yy_1 \sin \yy_1+\cos \yy_1)}{\yy_1} d\yy_1\right)\Bigg)\label{ed1}\\
d_2&=&\frac{1}{32 x_R^3}\Bigg( \left(\int_{\yy_i}^{2 x_R} \frac{12 \Pi^\aa (\yy_1) (\sin \yy_1-\yy_1 \cos \yy_1)}{\yy_1} d\yy_1\right)\Big(8 x_R^2 \sin x_R+4 x_R \cos x_R\nonumber\\
&&-\sin x_R+\sin 3 x_R\Big)+\left(8 x_R^2 \cos x_R-4 x_R \sin x_R-\cos x_R+\cos 3 x_R\right)\nonumber\\
&& \left(\int_{\yy_i}^{2 x_R} \frac{12 \Pi^\aa (\yy_1) (\yy_1 \sin \yy_1+\cos \yy_1)}{\yy_1} d\yy_1\right)\Bigg)\label{ed2}
\end{eqnarray}
Since we are interested in the production of GW from the EM field anisotropic stress, we neglect the homogeneous part in the $h^{\aa}$ expression for matter dominance given in Eq.~\eqref{hijinmd} while calculating the above expressions. Using Eq.~\eqref{maiequation}, we get the following expression for $d h^{\aa}/d x$,
\begin{align}
\frac{d h^{\aa}(\vec{k},x)}{d x}=&d_1 \left(\frac{\cos x}{x}-\frac{\sin x}{x^2}\right)+d_2 \left(\frac{\sin x}{x}-\frac{\cos x}{x^2}\right)-4 \left(\frac{\cos x}{x}-\frac{\sin x}{x^2}\right) \int_{x_R}^x d x_1 \Pi^{\aa}(\vec{k}, x_1) \nonumber\\
&y_0(x_1)+4 \left(\frac{\sin x}{x}-\frac{\cos x}{x^2}\right) \int_{x_R}^x d x_1 \Pi^{\aa}(\vec{k},x_1) j_0(x_1)
\end{align}
To estimate the GW spectrum, we need to calculate $|d h^\aa/dx|^2$. After multiplying the above expression with its complex conjugate and taking the limit $x >> 1$ (since we are interested in the modes which are deep inside the Hubble radius at the present epoch), we get,
\begin{align}
\Big|\frac{d h^{\aa}}{d x}\Big|^2(k,x)=&|d_1|^2 \left(\frac{\sin x}{x}\right)^2+|d_2|^2 \left(\frac{\cos x}{x}\right)^2+(d_1 d_2^*+d_1^*d_2)\frac{\sin x \cos x}{x^2}\nonumber\\
&+\frac{8}{x^2} \int_{x_R}^x d x_1 \int_{x_R}^x d x_2 |\Pi^{\aa}|^2(k, x_1,x_2) \frac{\cos(x_2-x_1)+\cos(2 x-x_1-x_2)}{x_1 x_2}
\end{align}
In the above expression, we only keep the terms proportional to $(1/x)^2$ and neglect the terms which have higher power of $1/x$ as those terms are subdominant in the limit $x>>1$. After reheating, the source term contributes until the epoch of neutrino decoupling. This is because anisotropic stress due to neutrinos comes into picture after neutrino decoupling and they balance the magnetic field anisotropic stress. Hence, there is no contribution to the GW energy density after neutrino decoupling. Further, averaging the above expression over a time scale greater than the time period of oscillation, we get,
\begin{align}\label{appendixeq}
\Big|\frac{d h^{\aa}}{d x}\Big|^2(k,x)=&\frac{1}{2 x^2}\left(|d_1|^2 +|d_2|^2\right)
+\frac{8}{x^2} \int_{x_R}^{x_{\nu d}}  \int_{x_R}^{x_{\nu d}} d x_1 d x_2 |\Pi^{\aa}|^2(k, x_1,x_2) \frac{\cos(x_2-x_1)}{x_1 x_2}
\end{align}
 
\section{Calculation for the estimate of GW energy spectrum}\label{3}
Here, we provide the calculation for analytical estimate of the GW energy spectrum for the modes $k< k_H$ given in Eq.~\eqref{analytical}. To evaluate GW energy spectrum, we estimate $|d h^{\aa}/d x|^2(k,x) $ given in Eq.~\eqref{appendixeq},
\begin{align}\label{hnsquare}
\Big|\frac{d h^{\aa}}{d x}\Big|^2(k,x)=&\frac{1}{2 x^2}\left(|d_1|^2 +|d_2|^2\right)
+\frac{8}{x^2} \int_{x_R}^{x_{\nu d}}  \int_{x_R}^{x_{\nu d}} d x_1 d x_2 |\Pi^{\aa}|^2(k, x_1,x_2) \frac{\cos(x_2-x_1)}{x_1 x_2}
\end{align}
In the above expresion, the first part in the R.H.S denotes the contribution before reheating and the second part denotes the contribution after reheating. In the post inflationary matter dominated era both electric and magnetic field anisotropic stresses contribute in the production of GW.  After reheating, only magnetic field anisotropic stresses contribute as electric fields get shorted out due to the large conductivity of the relativistic plasma. 
\subsection{Calculation for the contribution before reheating}
For $k<k_H=k_0/\beta$, $x_R= k \eta_R<1$. In this limit, the expression for $d_1$ and $d_2$ given in Eq.~\eqref{ed1} and Eq.~\eqref{ed2} respectively, reduces to,
\begin{align}
d_1=&\frac{3}{8 x_R^3}\Bigg(2 \int_{\yy_i}^{2 x_R} \frac{\yy_1^2}{3} \Pi^\aa (\yy_1)  d\yy_1+\frac{32 x_R^3}{3} \int_{\yy_i}^{2 x_R} \frac{1}{\yy_1} \Pi^\aa (\yy_1) d\yy_1\Bigg)\nonumber\\
d_2=&\frac{3}{8 x_R^3}\Bigg(-\frac{32 x_R^6}{45} \int_{\yy_i}^{2 x_R} \frac{1}{\yy_1} \Pi^\aa (\yy_1) d\yy_1+6 x_R \int_{\yy_i}^{2 x_R} \frac{\yy_1^2}{3} \Pi^\aa (\yy_1)  d\yy_1\Bigg)
\end{align}
Further the expression for $|d_1|^2$ and $|d_2|^2$ is given by,
\begin{align}
|d_1|^2=&\frac{9}{64 x_R^6}\Bigg( \int_{\yy_i}^{2 x_R} \int_{\yy_i}^{2 x_R} \left(\frac{4 \yy_1^2 \yy_2^2}{9}+\frac{1024 x_R^6}{9 \yy_1 \yy_2}+\frac{64 x_R^3(\frac{\yy_1^2}{\yy_2}+\frac{\yy_2^2}{\yy_1})}{9}\right)  |\Pi^\aa|^2(k,\yy_1,\yy_2) d\yy_1  d\yy_2\Bigg)\nonumber\\
|d_2|^2=&\frac{9}{64 x_R^6}\Bigg(36 x_R^2 \int_{\yy_i}^{2 x_R} \int_{\yy_i}^{2 x_R} \frac{\yy_1^2 \yy_2^2}{9} |\Pi^\aa|^2(k,\yy_1,\yy_2) d\yy_1  d\yy_2\Bigg)\nonumber
\end{align}
In the expression of $|d_2|^2$, we neglect the contribution from the first term in the expression of $d_2$ since this term is much smaller than the second term in the limit $x_R<1$. After substituting $|\Pi^\aa|^2$ from Eq.~\eqref{piinnonhelical}, we get
\begin{align}
|d_1|^2
=&\frac{9}{8 x_R^2}\left(\frac{1}{\tilde{\rho}+\tilde{p}}\right)^2\Bigg( \int_{\yy_i}^{2 x_R} \int_{\yy_i}^{2 x_R} \left(\frac{4 \yy_1^2 \yy_2^2}{9}+\frac{1024 x_R^6}{9 \yy_1 \yy_2}+\frac{64 x_R^3(\frac{\yy_1^2}{\yy_2}+\frac{\yy_2^2}{\yy_1})}{9}\right)  \left(\frac{1}{\yy_1^2 \yy_2^2}\right) \nonumber\\
&\Big(f_B(k,\yy_1,\yy_2)+f_E(k,\yy_1,\yy_2)\Big) d\yy_1  d\yy_2\Bigg)\label{d1square}
\end{align}
To evaluate the above expression, we first calculate $f_B(k,\yy_1,\yy_2)+f_E(k,\yy_1,\yy_2)$. Substituting Eq.~\eqref{defpowerspec} in Eq.~\eqref{fbfe}, we get,
\begin{align}
f_B(k,\yy_1,\yy_2)+f_E(k,\yy_1,\yy_2)=& \pi^2 \int_0^{\infty} \frac {d q }{q} \int_{-1}^{1} d \gamma \Bigg[\frac{d \tilde{\rho}_B(q, \yy_1)}{d \ln q} \frac{d \tilde{\rho}_B(|\vec{k}-\vec{q}|, \yy_1)}{d \ln |\vec{k}-\vec{q}|} C_{B}(q,\yy_1,\yy_2)\nonumber\\ 
&C_{B}(|\vec{k}-\vec{q}|,\yy_1,\yy_2)+\frac{d \tilde{\rho}_E(q, \yy_1)}{d \ln q} \frac{d \tilde{\rho}_E(|\vec{k}-\vec{q}|, \yy_1)}{d \ln |\vec{k}-\vec{q}|} C_{E}(q,\yy_1,\yy_2)\nonumber\\
& C_{E}(|\vec{k}-\vec{q}|,\yy_1,\yy_2)\Bigg] (1+\gamma^2+\beta^2+\gamma^2 \beta^2)
\end{align} 
It is evident from Eq.~\eqref{ms} and Eq.~\eqref{es} that the magnetic and electric spectral energy densities decay rapidly for $k\geq \kp(\eta)$. Keeping this in mind, we neglect the contribution of magnetic and electric power spectra for $k\geq \kp(\eta)$. Further, we take the upper limit of the $q$ integration to be $k_0$ instead of $\kp(\eta)$ because within one Hubble expansion time electric and magnetic field spectral energy densities increase by a very large value. The most of the contribution in the above integral is near the epoch of reheating. Hence, taking $k_0$ instead $\kp$ will not change our result much. Using Eq.~\eqref{ms} and Eq.~\eqref{es}, the above expression reduces to,
\begin{align}
f_B(k,\yy_1,\yy_2)+f_E(k,\yy_1,\yy_2)=& \pi^2 \int_0^{k_0} \frac {d q }{q} \int_{-1}^{1} d \gamma \Bigg[ D_1^2 \left(\frac{q}{k_0}\right)^4\frac{(k^2+q^2-2 k q \gamma)^{1/2}}{k_0^4}\left(\frac{\yy_1 \yy_2}{ (2 x_R)^2}\right)^{8 \beta+2}\nonumber\\
&+D_2^2\left(\frac{q}{k_0}\right)^2 \frac{(k^2+q^2-2 k q \gamma)^{-1/2}}{k_0^2} \left(\frac{\yy_1 \yy_2}{ (2 x_R)^2}\right)^{8 \beta}\Bigg] \nonumber\\
&(1+\gamma^2+\beta^2+\gamma^2 \beta^2)
\end{align} 

To solve the above expression, we first calculate the $\gamma$ integral as,
\begin{align}
\int_{-1}^{1} d \gamma  (k^2+q^2-2 k q \gamma)^{1/2}(1+\gamma^2+\beta^2+\gamma^2 \beta^2)&=\left\{
        \begin{array}{ll}
           \frac{8 \left(5 k^4+72 k^2 q^2+147 q^4\right)}{315 q^3}, & k\leq q\\
           \frac{16 \left(105 k^6+6 k^2 q^4+q^6\right)}{315 k^5},&k\geq q
        \end{array}
    \right. \nonumber\\
\int_{-1}^{1} d \gamma  (k^2+q^2-2 k q \gamma)^{-1/2}(1+\gamma^2+\beta^2+\gamma^2 \beta^2)&=\left\{
        \begin{array}{ll}
          \frac{8 \left(k^2+49 q^2\right)}{105 q^3},&  k\leq q\\
           \frac{16 \left(35k^4-72 k^2 q^2-3 q^4\right)}{105 k^5},&k\geq q
        \end{array}
    \right. \nonumber
    \end{align}
Further, we divide the $q$ integral into two parts,
$
\int_0^{k_0} dq=\int_0^{k}dq + \int_k^{k_0} dq$ and evaluate each of the part separately. We get,
\begin{align}
\int_0^{k_0} \frac{d q}{q} \int_{-1}^{1} d \gamma  \left(\frac{q}{k_0}\right)^4   \frac{(k^2+q^2-2 k q \gamma)^{1/2}}{k_0^4}(1+\gamma^2+\beta^2+\gamma^2 \beta^2)=&\frac{8}{1575 k_0^8} \Big(25 k^4 k_0+120 k^2 k_0^3\nonumber\\
&+147 k_0^5-21 k^5\Big)\nonumber\\
\int_0^{k_0} \frac{d q}{q} \int_{-1}^{1} d \gamma  \left(\frac{q}{k_0}\right)^2   \frac{(k^2+q^2-2 k q \gamma)^{-1/2}}{k_0^2}(1+\gamma^2+\beta^2+\gamma^2 \beta^2)&=\frac{4 \left(98k_0^2-2 k^2-35k k_0\right)}{105 k_0^5}
\end{align}
After taking only the contribution from the dominating terms, we get,
\begin{align}
f_B(k,\yy_1,\yy_2)+f_E(k,\yy_1,\yy_2)=& \pi^2 \Bigg[D_1^2 \frac{56}{75 k_0^3}\left(\frac{\yy_1 \yy_2}{ (2 x_R)^2}\right)^{8 \beta+2}+D_2^2\frac{56}{15 k_0^3} \left(\frac{\yy_1 \yy_2}{ (2 x_R)^2}\right)^{8 \beta}\Bigg] 
\end{align} 
Since during the post inflationary matter dominated era, the electric spectral energy density dominates over the magnetic spectral energy density, $D_1< D_2$, the quantity $\left(\frac{\yy_1 \yy_2}{ (2 x_R)^2}\right)$ is always less than unity. Therefore we neglect the first term in comparison to the second term in the above expression. This implies,
\begin{align}
f_B(k,\yy_1,\yy_2)+f_E(k,\yy_1,\yy_2)=& \pi^2 \Bigg[D_2^2\frac{56}{15 k_0^3} \left(\frac{\yy_1 \yy_2}{ (2 x_R)^2}\right)^{8 \beta}\Bigg] 
\end{align}
After substituting the above expression in Eq.~\eqref{d1square} and using new variables for integration defined as $z_1=\yy_1/(2 x_R)$ and $z_2=\yy_2/(2 x_R)$, we get
\begin{align}
|d_1|^2=&\frac{9  \pi^2 D_2^2}{2}\frac{56}{15 k_0^3}\left(\frac{1}{\tilde{\rho}+\tilde{p}}\right)^2\Bigg( \int_{z_i}^{1} \int_{z_i}^{1} \left(\frac{4}{9}+\frac{16}{9 z_1^3 z_2^3}+\frac{8}{9} \left(\frac{1}{z_1^3}+\frac{1}{z_2^3}\right)\right)  \left(z_1 z_2\right)^{8 \beta} d z_1 d z_2 
\end{align}
After calculating the above integral we get,
\begin{align}
|d_1|^2=&\frac{9  \pi^2 D_2^2}{2}\frac{56}{15 k_0^3}\left(\frac{1}{\tilde{\rho}+\tilde{p}}\right)^2\frac{64 \beta^2}{(1-4 \beta )^2 (8 \beta +1)^2} 
\end{align}
Similarly we can calculate the expression for $|d_2|^2$,
\begin{align}
|d_2|^2=&\frac{9  \pi^2 D_2^2}{2}\frac{56}{15 k_0^3}\left(\frac{1}{\tilde{\rho}+\tilde{p}}\right)^2\frac{4 x_R^2}{ (8 \beta +1)^2}.
\end{align}
From the above expression of $|d_1|^2$ and $|d_2|^2$, we see that $|d_1|^2$ is larger than $|d_2|^2$ for $x_R<1$. Therefore we neglect the contribution from $|d_2|^2$ in our further calculation.
After substituting $|d_1|^2$ in Eq.~\eqref{hnsquare}, we get 
\begin{align}
\Big|\frac{d h^{\aa}}{d x}\Big|^2(k,x)&=\frac{42 \pi^2}{5 x^2} \left(\frac{1}{k_0}\right)^3 \left(\frac{D_2}{\tilde{\rho}+\tilde{p}}\right)^2\left(\frac{64 \beta^2}{(1-4 \beta )^2 (8 \beta +1)^2}\right).
\end{align}
Further substituting the above expression in Eq.~\eqref{omegainhp}, we obtain the following expression for the GW energy spectrum,
\begin{align}
\frac{d\Omega_{GW}}{d \ln k}\Bigg|_0&=\frac{7 \Omega_R}{5} \left(\frac{k}{k_0}\right)^3 \left(\frac{D_2}{\tilde{\rho}+\tilde{p}}\right)^2\left(\frac{64 \beta^2}{(1-4 \beta )^2 (8 \beta +1)^2}\right).
\end{align}

\subsection{Calculation for the contribution after reheating}\label{4}
To evaluate GW energy spectrum after reheating, we estimate $|d h^{\aa}/d x|^2(k,x) $ given in Eq.~\eqref{hnsquare},
\begin{align}
\Big|\frac{d h^{\aa}}{d x}\Big|^2_{x>x_R}(k,x)=&\frac{8}{x^2} \int_{x_R}^{x_{\nu d}}  \int_{x_R}^{x_{\nu d}} d x_1 d x_2 |\Pi^{\aa}|^2(k, x_1,x_2) \frac{\cos(x_2-x_1)}{x_1 x_2}
\end{align}
After substituting $|\Pi^\aa|^2$ from Eq.~\eqref{piinnonhelical}, we get
\begin{align}
\Big|\frac{d h^{\aa}}{d x}\Big|^2_{x>x_R}(k,x)
=&\frac{4}{x^2} \left(\frac{1}{\tilde{\rho}+\tilde{p}}\right)^2 \int_{x_R}^{x_{\nu d}}  \int_{x_R}^{x_{\nu d}} d x_1 d x_2 f_B(k,\yy_1,\yy_2) \frac{\cos(x_2-x_1)}{x_1 x_2}\label{c15}
\end{align}
To evaluate the above expression, we first calculate $f_B(k,\yy_1,\yy_2)$. Substituting Eq.~\eqref{defpowerspec} in Eq.~\eqref{fbfe}, we get,
\begin{align}
f_B(k,x_1,x_2)=& \pi^2 \int_0^{\infty} \frac {d q }{q} \int_{-1}^{1} d \gamma \frac{d \tilde{\rho}_B(q, x_1)}{d \ln q} \frac{1}{(|\vec{k}-\vec{q}|)^3}\frac{d \tilde{\rho}_B(|\vec{k}-\vec{q}|, x_1)}{d \ln |\vec{k}-\vec{q}|} C_{B}(q,x_1,x_2)C_{B}(|\vec{k}-\vec{q}|,x_1,x_2) (1+\gamma^2+\beta^2+\gamma^2 \beta^2)\label{c16}
\end{align} 
Using Eqs.~\eqref{c15}, \eqref{c16} and \eqref{omegainhp}, we get
\begin{align}
  \frac{d\Omega_{GW}}{d \ln k}\Bigg|_0=&\frac{2~ \Omega_R}{3 (\tilde{\rho}+\tilde{p})^2} k^3  \int_{x_R}^{x_{\nu d}}  \int_{x_R}^{x_{\nu d}} d x_1 d x_2  \frac{\cos(x_2-x_1)}{x_1 x_2} \int_0^{\infty} \frac {d q }{q} \int_{-1}^{1} d \gamma \frac{d \tilde{\rho}_B(q, x_1)}{d \ln q} \frac{1}{(|\vec{k}-\vec{q}|)^3}\frac{d \tilde{\rho}_B(|\vec{k}-\vec{q}|, x_1)}{d \ln |\vec{k}-\vec{q}|} \nonumber\\
  &C_{B}(q,x_1,x_2)C_{B}(|\vec{k}-\vec{q}|,x_1,x_2) (1+\gamma^2+\beta^2+\gamma^2 \beta^2)
\end{align}
The magnetic energy spectrum peaks at $k=k_{NL}$, the main contribution to the integral comes when $q \sim k_{NL}$ and $|\vec{k}-\vec{q}|\sim k_{NL}$. For the case $k<<k_{NL}$, $|\vec{k}-\vec{q}| \sim q$ and $C_B$ changes from 1 to a small value within one Hubble time. Therefore the dominant contribution comes within one Hubble time from reheating.
\begin{align}
  \frac{d\Omega_{GW}}{d \ln k}\Bigg|_0\approx&\frac{2~ \Omega_R}{3 (\tilde{\rho}+\tilde{p})^2} k^3  \int_{x_R}^{2 x_{R}}  \int_{x_R}^{2 x_{R}} d x_1 d x_2  \frac{\cos(x_2-x_1)}{x_1 x_2}\left[ \frac{D_1^2}{k_{NL}^3}\left(\frac{x_1}{x_R}\right)^{-4/3}\left(\frac{x_1}{x_R}\right)^{-4/3} \right]C_{B}(k_{NL},x_1,x_2)^2 \nonumber\\
  &\int_{-1}^{1} d \gamma (1+\gamma^2+\gamma^2+\gamma^4)\nonumber\\
  =&\frac{56}{15}\frac{2~ \Omega_R}{3}\left(\frac{D_1}{\tilde{\rho}+\tilde{p}}\right)^2  \left(\frac{k}{k_0}\right)^3 \int_{x_R}^{2 x_{R}}  \int_{x_R}^{2 x_{R}} d x_1 d x_2  \frac{\cos(x_2-x_1)}{x_1 x_2}\left(\frac{x_1}{x_R}\right)^{-4/3}\left(\frac{x_1}{x_R}\right)^{-4/3}C_{B}(k_{NL},x_1,x_2)^2\nonumber
  \end{align}
  In terms of the variable $z_1=x_1/x_R$ and $z_2=x_2/x_R$, the above expression reduces to,
  \begin{align}
  \frac{d\Omega_{GW}}{d \ln k}\Bigg|_0\approx&\frac{56}{15}\frac{2~ \Omega_R}{3}\left(\frac{D_1}{\tilde{\rho}+\tilde{p}}\right)^2  \left(\frac{k}{k_0}\right)^3 \int_{1}^{2}  \int_{1}^{2} d z_1 d z_2  \frac{\cos(x_R^2(z_2-z_1))}{z_1 z_2}\left(z_1\right)^{-8/3}\left(z_1\right)C_{B}(k_{NL},z_1,z_2)^2\nonumber
  \end{align}
  The modes $k<k_H$ are outside the Hubble horizon at reheating. For these modes, $x_R<1$ so we can approximate $\cos(x_R^2(z_2-z_1))\sim 1$ for these modes. Using this, the GW spectrum for the modes $k<k_H$, we get,
   \begin{align}
  \frac{d\Omega_{GW}}{d \ln k}\Bigg|_0\approx&\frac{56}{15}\frac{2~ \Omega_R}{3}\left(\frac{D_1}{\tilde{\rho}+\tilde{p}}\right)^2  \left(\frac{k}{k_0}\right)^3 \int_{1}^{2}  \int_{1}^{2} \frac{d z_1 d z_2}{z_1 z_2} z_1^{-8/3} z_1 C_{B}(k_{NL},z_1,z_2)^2\nonumber\\
  =&c~\Omega_R \left(\frac{D_1}{\tilde{\rho}+\tilde{p}}\right)^2 \left(\frac{k}{k_0}\right)^3
  \end{align}
  where,
  \begin{align*}
      c=\frac{56}{15}\frac{2}{3} \int_{1}^{2}  \int_{1}^{2} \frac{d z_1 d z_2}{z_1 z_2}\left(z_1\right)^{-8/3} z_1 C_{B}(k_{NL},z_1,z_2)^2
  \end{align*}
which has different value for different $k_{NL}$ and $\epsilon$ since unequal time correlation function, $C_B$ depends upon $k_{NL}$ and $\epsilon$. For $T_R=100$ GeV and $\epsilon=1$, $c=0.18$.
As is evident from the above expression, GW energy spectrum is proportional to $k^3$ and to the fraction of magnetic field energy density to the background energy density for the modes $k<k_H$.
\bibliographystyle{apsrev4-1}
\bibliography{references}
\end{document}